\documentclass[traditabstract]{aa}

\usepackage{epsfig}    
\usepackage{lscape}
\usepackage{natbib}
\bibpunct{(}{)}{;}{a}{}{,}    
\usepackage{graphicx,graphics}
\usepackage{color} 
\usepackage{times}
\usepackage{amsmath}
\usepackage{amssymb}
\newcommand{\MC}{\multicolumn}
\begin{document}

\title{Oxygen abundance distributions in six late-type galaxies 
       based on SALT spectra of  H\,{\sc ii} regions\footnotemark[0]\thanks{Based
on observations made with the Southern African Large 
Telescope, programs \mbox{2012-1-RSA\_OTH-001}, \mbox{2012-2-RSA\_OTH-003} and
\mbox{2013-1-RSA\_OTH-005}.}}

\author{
        I.~A.~Zinchenko\inst{\ref{ARI},\ref{MAO}} \and
	A.~Y.~Kniazev\inst{\ref{SAAO},\ref{SALT},\ref{SAI}} \and
        E.~K.~Grebel\inst{\ref{ARI}} \and
        L.~S.~Pilyugin\inst{\ref{ARI},\ref{MAO},\ref{KFU}}
       }
       
\institute{Astronomisches Rechen-Institut, Zentrum f\"{u}r Astronomie 
der Universit\"{a}t Heidelberg, 
M\"{o}nchhofstr.\ 12--14, 69120 Heidelberg, Germany \label{ARI} 
\and
Main Astronomical Observatory, National Academy of Sciences of Ukraine, 
27 Akademika Zabolotnoho St., 03680, Kyiv, Ukraine \label{MAO}
\and
South African Astronomical Observatory, PO Box 9, 7935 Observatory,
Cape Town, South Africa \label{SAAO}
\and
Southern African Large Telescope Foundation, PO Box 9, 7935 Observatory,
Cape Town, South Africa \label{SALT}
\and
Sternberg Astronomical Institute, Lomonosov Moscow State University,
Universitetskij Pr.\ 13, Moscow 119992, Russia \label{SAI}
\and
Kazan Federal University, 18 Kremlyovskaya St., 420008, Kazan, Russian Federation \label{KFU}
}

\abstract{
Spectra of 34 H\,{\sc ii} regions in the late-type galaxies NGC~1087,
NGC~2967, NGC~3023, NGC~4030, NGC~4123, and NGC~4517A were observed
with the  South African Large Telescope (SALT). In all 34 H\,{\sc ii} 
regions, oxygen abundances were determined through the ``counterpart'' 
method ($C$ method).  Additionally, in two  H\,{\sc ii} regions in
which the auroral lines were detected oxygen abundances were measured
through the classic $T_{e}$ method.  We also estimated the 
abundances in our H\,{\sc ii} regions using the O3N2 and N2 calibrations 
and compared those with the $C$-based abundances.  
With these data we examined the radial abundance distributions in 
the disks of our target galaxies.  We derived surface-brightness 
profiles and other characteristics of the disks (the surface brightness 
at the disk center and the disk scale length) in three photometric bands 
for each galaxy using publicly available photometric imaging data. The radial 
distributions of the oxygen abundances predicted by the relation between 
abundance and disk surface brightness in the $W1$ band obtained for 
spiral galaxies in our previous study are close to the radial 
distributions of the oxygen abundances determined from the analysis 
of the emission line spectra for four galaxies where this relation is 
applicable.  Hence, when the surface-brightness profile of a late-type
galaxy is known, this parametric relation can be used to estimate the
likely present-day oxygen abundance in its disk. 
}


\keywords{galaxies: abundances -- ISM: abundances 
-- H\,{\sc ii} regions, galaxies: individual: 
NGC~1087, NGC~2967, NGC~3023, NGC~4030, NGC~4123, NGC~4517A}

\titlerunning{Oxygen abundance distributions in six galaxies}
\authorrunning{Zinchenko et al.}
\maketitle

\section{Introduction}

Metallicities play a key role in studies of galaxies.  The present-day
abundance distributions across a galaxy provide important information
about the evolutionary status of that galaxy and form the basis for
the construction of models of the chemical evolution of galaxies.  

Oxygen abundances and their gradients in the disks of late-type
galaxies are typically based on emission-line spectra of individual
H\,{\sc ii} regions.  When the auroral line [O\,{\sc
iii}]$\lambda$4363 is detected in the spectrum of an H\,{\sc ii}
region, the $T_{e}$-based oxygen (O/H)$_{T_{e}}$ abundance can be
derived using the standard equations of the $T_{e}$-method.  In our
current study, we do not always have this information and use
alternative methods where the auroral line is not detected.  In those
cases, we estimate the oxygen abundances from strong emission lines
using a recently suggested method (called the ``$C$ method'') for
abundance determinations of \citet{Pilyugin2012} and
\citet{Pilyugin2014a}.  When the strong lines $R_3$, $N_2$, and $S_2$
are measured in the spectrum of an H\,{\sc ii} region, the oxygen
(O/H)$_{C_{\rm SN}}$ abundance can be determined.  When the strong
lines $R_2$, $R_3$, and $N_2$ are measured in the spectrum, we can
measure the oxygen (O/H)$_{C_{\rm ON}}$ abundance.

It should be emphasized that the $C$ method produces abundances on the
same metallicity scale as the $T_{e}$-method.  In contrast,
metallicities derived using one of the many calibrations based on
photoionization models tend to show large discrepancies (of up to
$\sim$0.6 dex) with respect to $T_{e}$-based abundances \citep[see the reviews
by][]{Kewley2008,LopezSanchezEsteban2010AA517,LopezSanchezetal2012MNRAS426}.
Coupling our emission-line measurements with available line
measurements from the literature or public databases, we measured the
radial distributions of the oxygen abundances across the disks of six
galaxies. 

The study of the correlations between the oxygen abundance and other
properties of spiral and irregular galaxies is important for 
understanding the formation and evolution of galaxies.  The
correlation between the local oxygen abundance and the stellar surface
brightness (the OH -- $SB$ relation) or surface mass density has been
a subject of discussion for a long time
\citep{Webster1983,Edmunds1984,VilaCostas1992,Ryder1995,Moran2012,Rosales2012,Sanchez2014}.
We examined the relations between the oxygen abundance and the disk
surface brightness in the infrared $W1$ band at different fractions of
the optical isophotal radius $R_{25}$ in our previous paper
\citep{Pilyugin2014b}. $W1$ is the photometric band of the {\it
Wide-field Infrared Survey Explorer (WISE)}; see \citet{Wright2010}.
We found evidence that the OH -- $SB$ relation depends on the
galactocentric distance (taken as a fraction of the optical radius
$R_{25}$) and on other properties of a galaxy, namely its disk scale
length and the morphological $T$-type.  In that study, we suggested a
parametric OH -- $SB$ relation for spiral galaxies. 

In our current paper, we present results from observations of
emission-line spectra of H\,{\sc ii} regions in six spiral galaxies.
These observations were obtained with the South African Large
Telescope as a part of our investigation of the abundance properties
of nearby late-type galaxies \citep{Pilyugin2014a,Pilyugin2014b}. 

We constructed radial surface-brightness profiles of our galaxies in
the infrared $W1$ band using the photometric maps obtained by the {\it
WISE} satellite \citep{Wright2010}.  The characteristics of the disk for
each galaxy were obtained through bulge-disk decomposition.  The
radial distributions of the oxygen abundances predicted by the
parametric OH -- $SB$ relation are compared to the radial
distributions of the oxygen abundances determined from the analysis of
the emission-line spectra of the H\,{\sc ii} regions in our target
galaxies. 

The paper is organized as follows.  Our sample of galaxies is
presented in Section 2.  The spectroscopic observations and data
reduction are described in Section 3.  The photometric properties of
our galaxies are discussed in Section 4.  The oxygen abundances are
presented in Section 5.  Section 6 contains a discussion and a brief
summary of the main results.  

Throughout the paper, we will use the following standard notations 
for the line intensities $I$: \\ 
$R$  = $I_{\rm [O\,III] \lambda 4363} /I_{{\rm H}\beta }$,  \\
$R_2$  = $I_{\rm [O\,II] \lambda 3727+ \lambda 3729} /I_{{\rm H}\beta }$,  \\
$N_2$  = $I_{\rm [N\,II] \lambda 6548+ \lambda 6584} /I_{{\rm H}\beta }$,  \\
$S_2$  = $I_{\rm [S\,II] \lambda 6717+ \lambda 6731} /I_{{\rm H}\beta }$,  \\
$R_3$  = $I_{{\rm [O\,III]} \lambda 4959+ \lambda 5007} /I_{{\rm H}\beta }$ .

\section{Our galaxy sample}

Our original sample of spiral galaxies for follow-up observations with 
the Southern African Large Telescope \citep[SALT;][]{Buck06,Dono06}
was devised based on Sloan Digital Sky Survey (SDSS) images and  
the fact that SALT can observe targets with a declination $\delta <10$~degrees
and has a field of view of 8~arcmin.
Each selected spiral galaxy contains a sufficiently large number of bright
\ion{H}{ii} regions distributed across the whole galaxy disk and fitting
SALT's field of view.
The total sample consists of $\sim30$ nearby galaxies
that are located in the equatorial sky region. 

\begin{figure*}
  \begin{center}
  \includegraphics[width=0.3\linewidth]{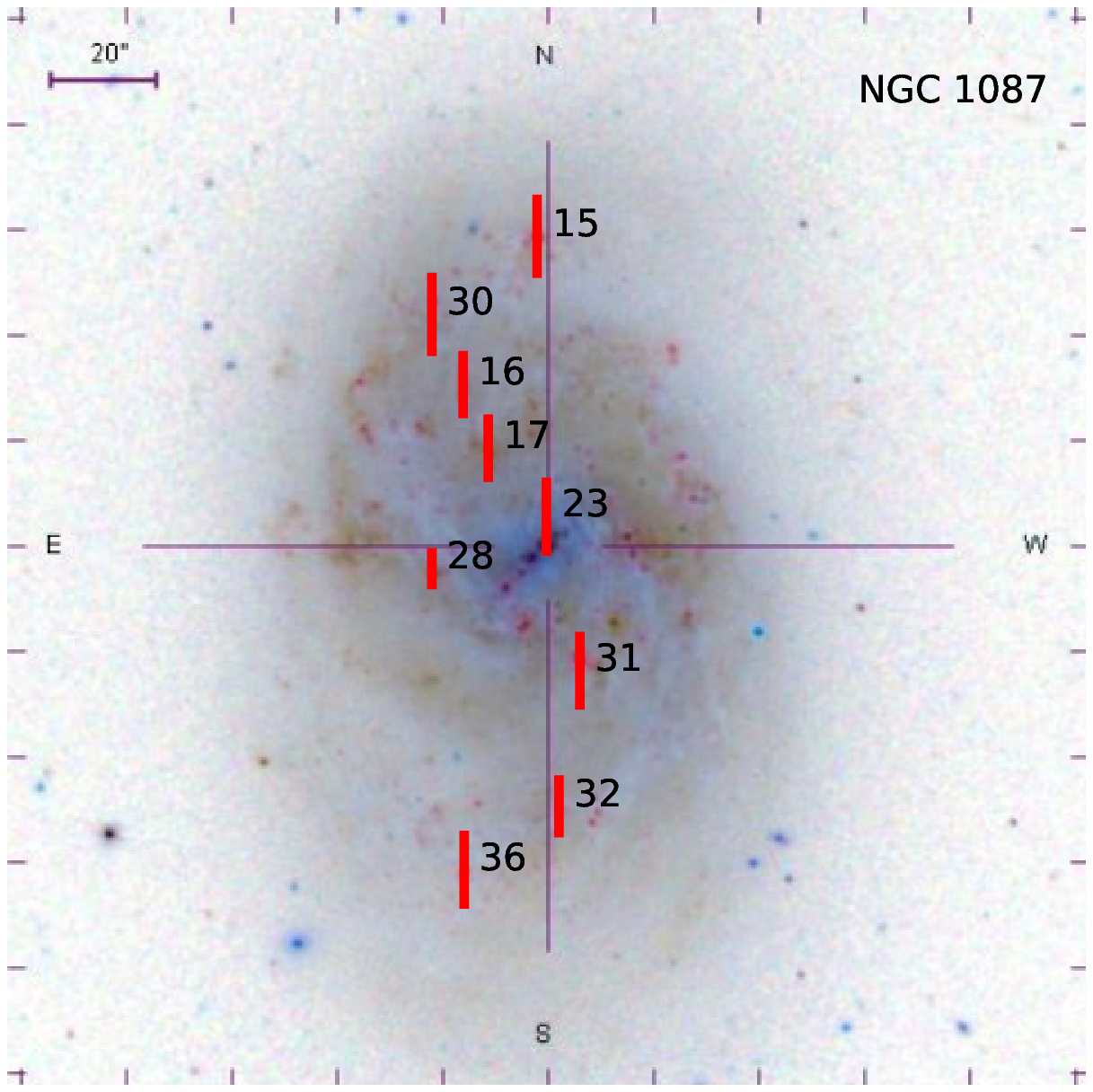}
  \includegraphics[width=0.3\linewidth]{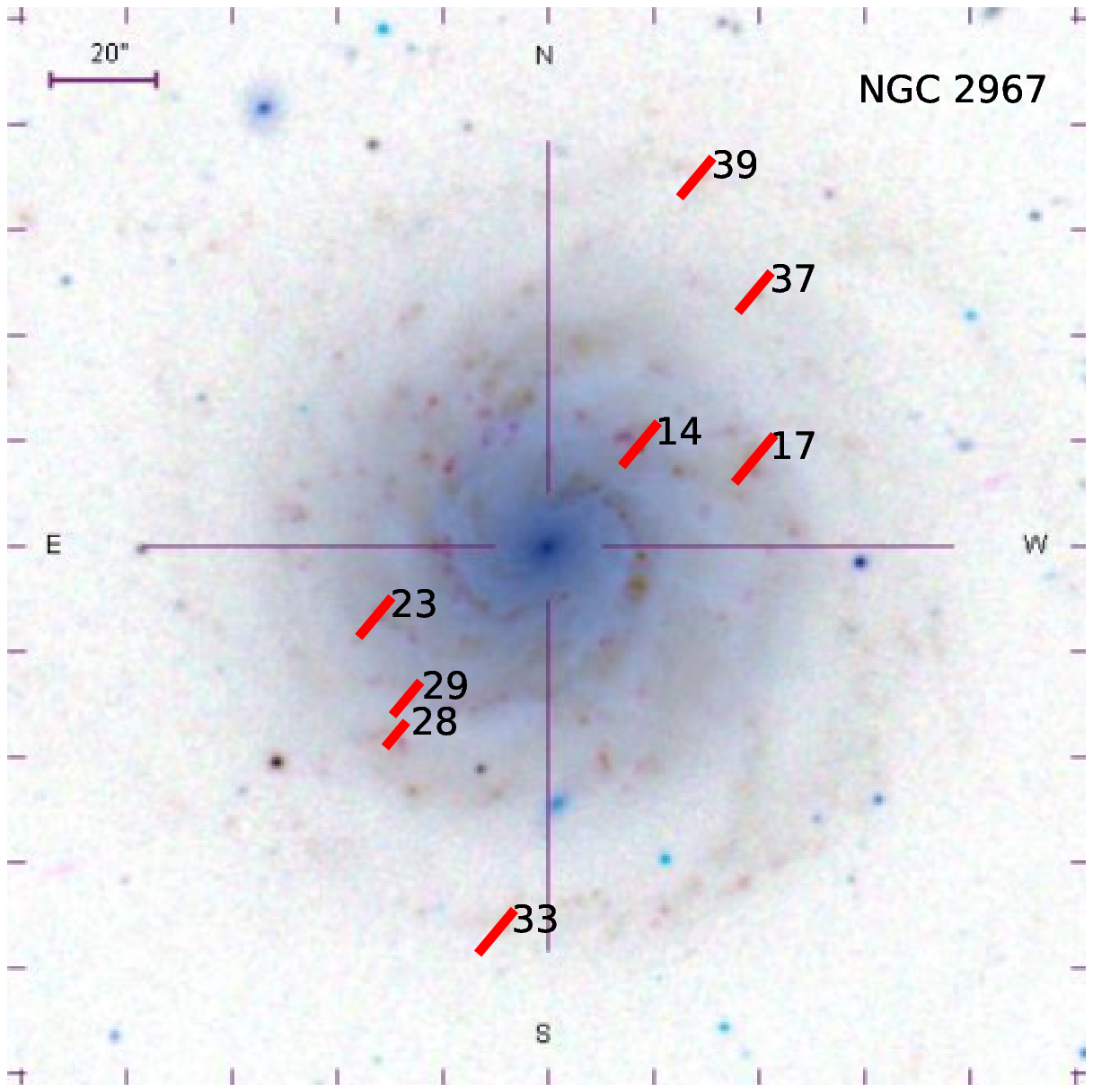}  
  \includegraphics[width=0.3\linewidth]{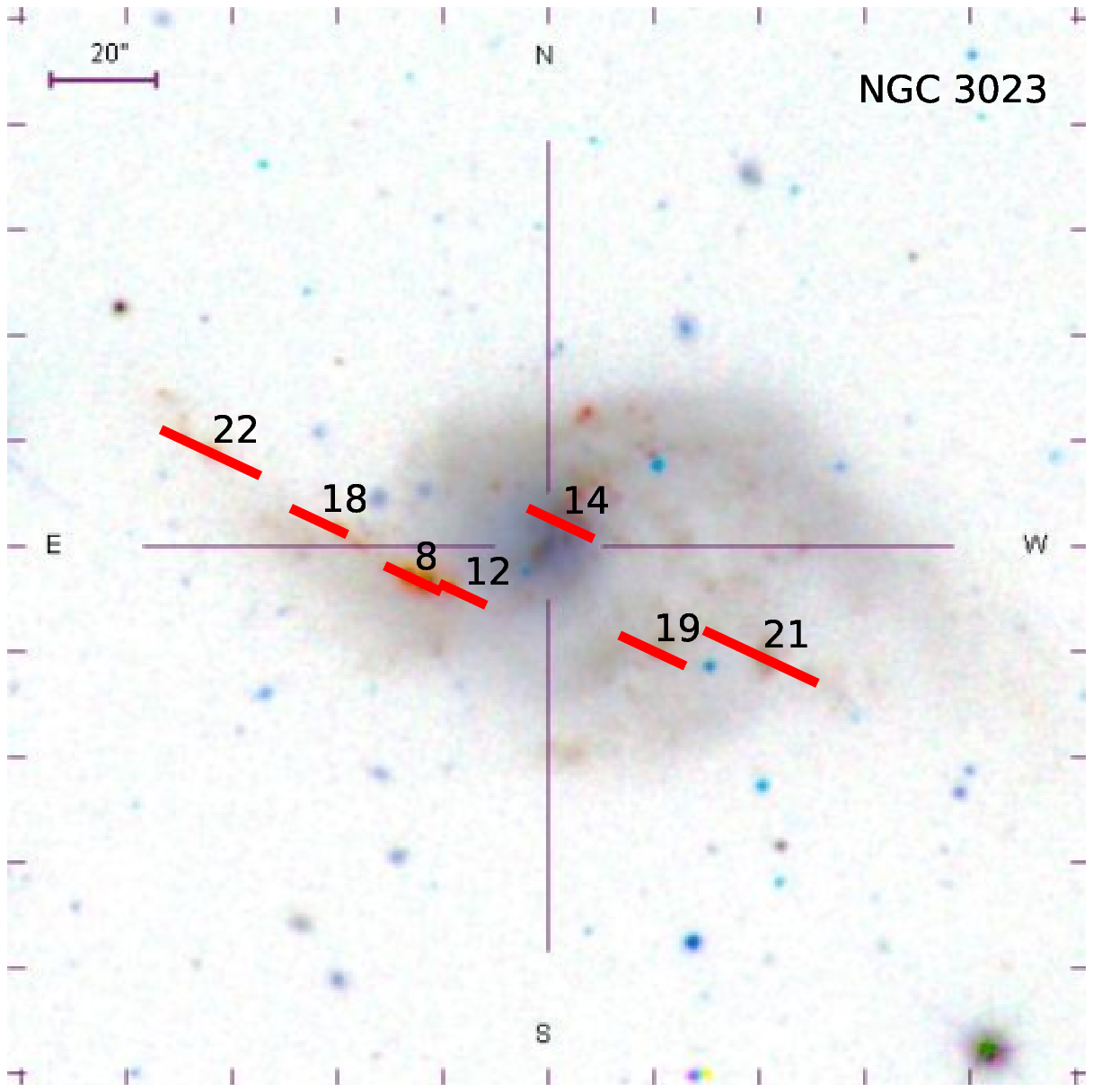}
  \includegraphics[width=0.3\linewidth]{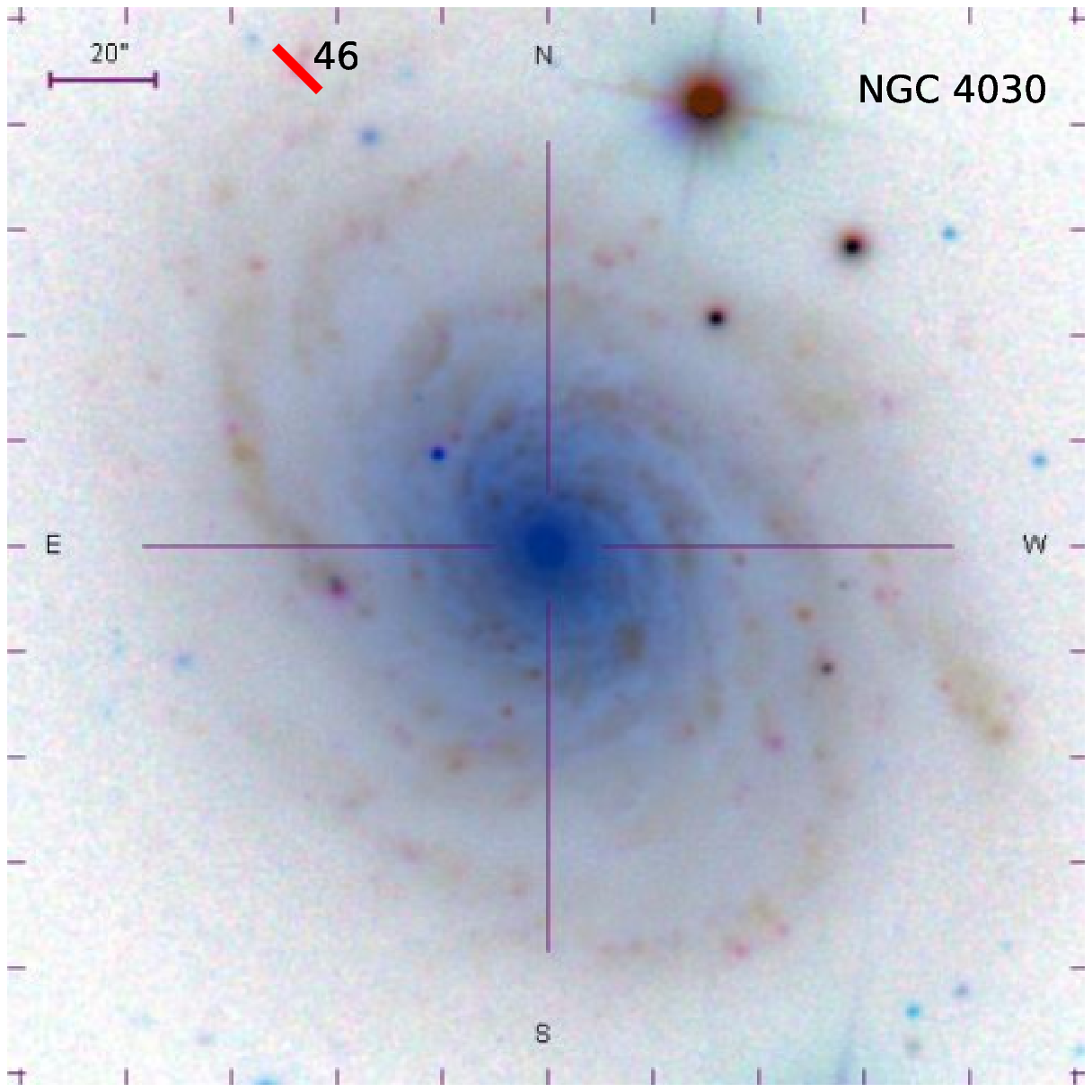}
  \includegraphics[width=0.3\linewidth]{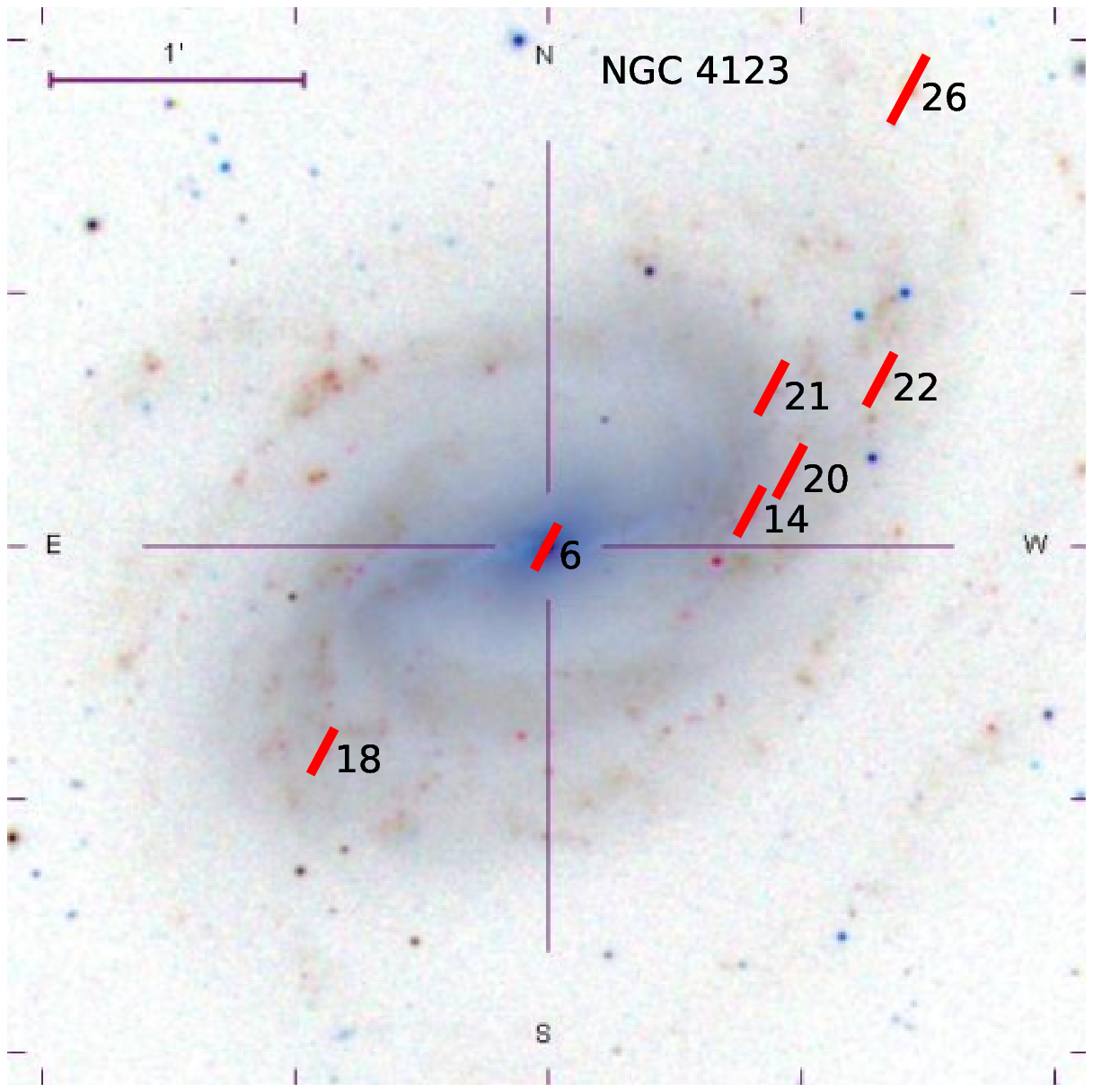}  
  \includegraphics[width=0.3\linewidth]{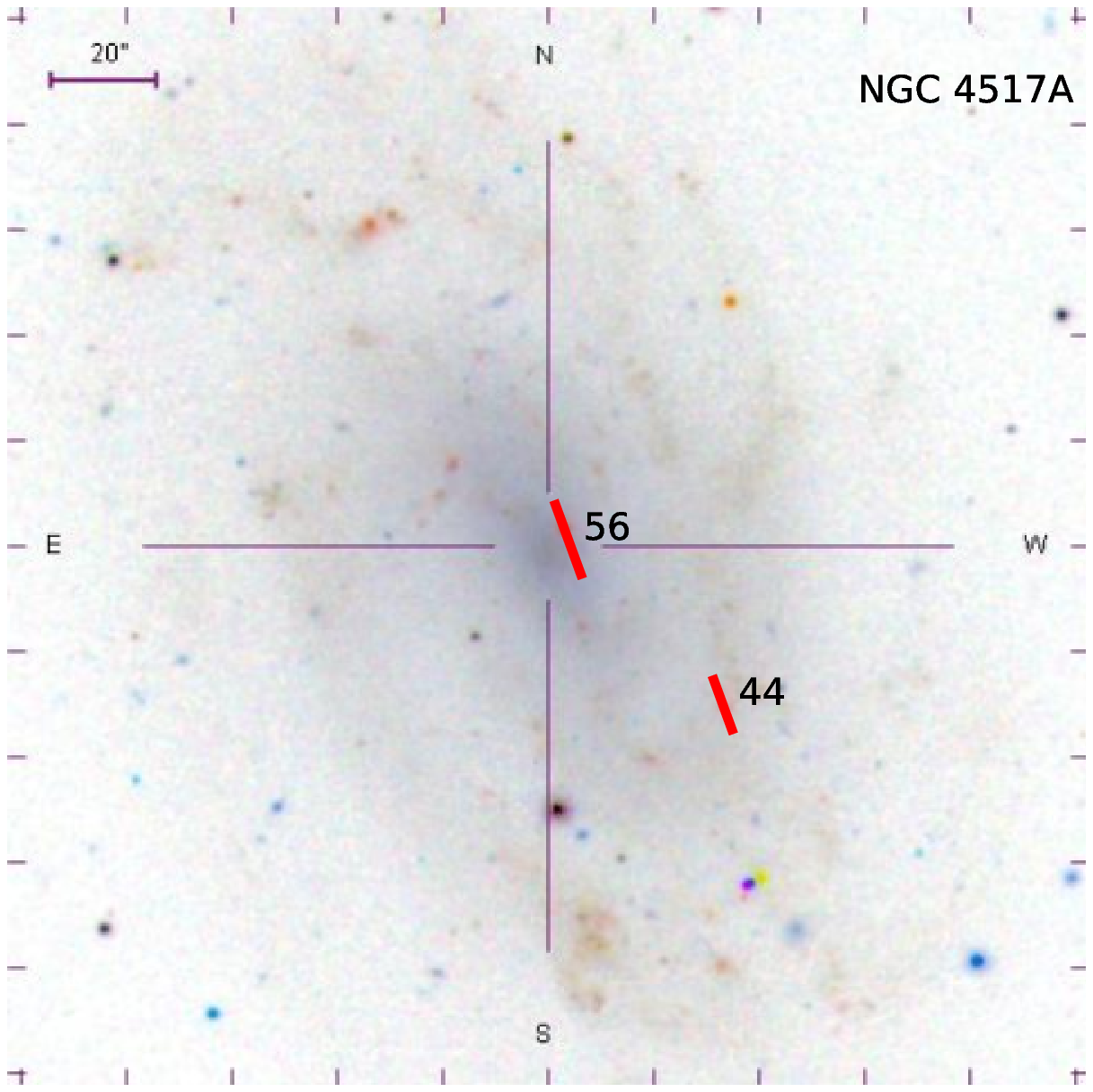}
  \end{center}
\caption{{\it SDSS} images of our target galaxies with marked positions of 
the slits used in the present investigation.
}
\label{figure:images}
\end{figure*}

Out of this sample of 58 galaxies, we have obtained spectra of 
H\,{\sc ii} regions in six galaxies (NGC~1087, 
NGC~2967, NGC~3023, NGC~4030, NGC~4123, NGC~4517A) thus far. 
In Fig.~\ref{figure:images} we present the images and slit positions 
for those galaxies.  For a detailed description of the observations
we refer to Section 3.

Table \ref{table:sample} lists the general characteristics of each
galaxy.  We listed the most widely used identifications for our
target galaxies, i.e., the designations in the New General Catalogue (NGC)
and in the Uppsala General Catalog of Galaxies (UGC).  The
morphological type of the galaxy and morphological type code $T$ were
adopted from {\sc leda} \citep[Lyon-Meudon Extragalactic
Database;][]{Paturel1989,Paturel2003}. The right ascension  and
declination were taken from the NASA/IPAC Extragalactic Database ({\sc
ned})\footnote{The NASA/IPAC Extragalactic Database ({\sc ned}) is
operated by the Jet Populsion Laboratory, California Institute of
Technology, under contract with the National Aeronautics and Space
Administration.  {\tt http://ned.ipac.caltech.edu/} }.  The
inclination of each galaxy, the position angle of the major axis, and
the isophotal radius $R_{25}$ in arcmin of each galaxy were determined
in our current study.  The distances were taken from {\sc ned}. These
distances include flow corrections for the Virgo cluster, the Great 
Attractor, and
Shapley Supercluster infall.  The isophotal radius in kpc was estimated
from the isophotal radius $R_{25}$ in arcmin and the distance listed
above.  The characteristics of the disks (the surface brightness at
the disk center in the $W1$ band and the disk scale length) were
determined through the bulge-disk decomposition carried out in
our current paper.  The surface brightness at the disk center was
reduced to a face-on galaxy orientation and is given in terms of
$L_{\sun}$ pc$^{-2}$.

\begin{table*}
\caption[]{\label{table:sample}
The adopted and derived properties of our target galaxies
}
\begin{center}
\begin{tabular}{lcccccc} \hline \hline
Name                                             & 
NGC~1087                                     & 
NGC~2967                                     & 
NGC~3023                                     & 
NGC~4030                                     & 
NGC~4123                                     & 
NGC~4517A                                    \\  
                                             & 
UGC~2245                                     & 
UGC~5180                                     & 
UGC~5269                                     & 
UGC~6993                                     & 
UGC~7116                                     & 
UGC~7685                                     \\  \hline
Morphological type, type code $T$                                    &  SABc,  5.0    &  Sc,   5.2     &   SABc, 5.5    &  Sbc,  4.0       &   Sc,   5.0     &   SBdm,   8.0    \\
Right ascension (J2000.0) [deg]                                                &   41.604852    &  145.513729    &  147.469112    &  180.098445      &  182.046296     &  188.117319      \\
Declination (J2000.0) [deg]                                          &   -0.498649    &    0.336438    &    0.618167    &   -1.100095      &    2.878283     &    0.389670      \\
Inclination [deg], ellipticity                                       &    53.0,  0.39 &  21.6, 0.07    &    49.5, 0.35  &    36.9, 0.20    &    43.1, 0.27   &     50.2, 0.36   \\
Position angle [deg]                                                 &     1          &  151           &    86          &    38            &    128          &     23           \\
Isophotal radius $R_{25}$$^{a}$   [arcmin]                           &     1.86       &   1.33         &     1.02       &     1.89         &     1.69        &      1.40        \\
Distance  [Mpc]                                                      &    20.1        &  29.5          &    29.4        &    26.4          &    27.3         &     27.8         \\
Isophotal radius $R_{25}$   [kpc]                                    &    10.86       &  11.41         &     8.72       &    14.51         &    13.42        &     11.32        \\
Disk scale length in $W1$ band      [kpc]                            &    2.45        &  1.97          &                &    3.36          &    4.70         &     4.90         \\
Logarithm of $W1$ brightness of disk center$^b$ [$L_{\odot}$ pc$^{-2}$] &    3.056       &  3.315         &                &    3.234         &    2.557        &     1.955        \\
\hline 
\end{tabular}\\
\end{center}
\begin{flushleft}
$^a$ in the $B$ band \\
$^b$ reduced to a face-on galaxy orientation  \\

\end{flushleft}
\end{table*}

Three galaxies of our sample are members of pairs of galaxies
and are included in the ``Catalogue of Isolated Pairs of Galaxies in
the Northern Hemisphere'' \citep{Karachentsev1972}: NGC~3023 = 
KPG~216B, NGC~4123 = KPG~322B, NGC~4517A = KPG~344A.  Two galaxies of our
sample, NGC~2967 and NGC~4030, are known members of galaxy groups
\citep{Fouque1992,Garcia1993}.  The galaxy NGC~4517A is a
low-surface-brightness galaxy \citep{Romanishin1983}.

\section{Spectroscopic observations and reduction}

\subsection{Observing procedures}

The spectroscopic observations of our selected galaxies were obtained
with the multi-object spectroscopic mode (MOS) of the Robert Stobie
Spectrograph  \citep[RSS;][]{Burgh03,Kobul03} installed at SALT.  
For each galaxy from our sample we constructed a MOS mask, where the slit
positions for the \ion{H}{ii} regions were selected using $gri$ SDSS images.
A typical MOS mask thus devised contained 9--12 slits for \ion{H}{ii} regions distributed over 
the whole galaxy disk. The usual width of the slits was 1.5 arcsec.  
The general strategy of the observations was to cover the total spectral range of 3600--7000~\AA\
in order to detect (1) the Balmer lines H$\delta$, H$\gamma$, H$\beta$ and H$\alpha$, which were used for extinction corrections, 
and (2) various emission lines used for the calculation of abundances: [\ion{O}{ii}] $\lambda$3727,3729,
[\ion{O}{iii}] $\lambda$4363, [\ion{O}{iii}] $\lambda$4959,5007, [\ion{N}{ii}] $\lambda$6548,6584 
and [\ion{S}{ii}] $\lambda$6717,6731.
We chose a resolution of $R=1000$--2000 in order
to be able to resolve emission lines located close to each other, e.g., H$\gamma$ and [\ion{O}{iii}] $\lambda$4363,
or [\ion{S}{ii}] $\lambda$6717 and [\ion{S}{ii}] $\lambda$6731.

In MOS mode the actual spectral coverage for each slit varies slightly depending on the
X-position of a given slit relative to the center of the mask.
Thus, to cover the total desired spectral range of 3600--7000~\AA\ for each slit of the mask 
we had to obtain two spectral setups for each studied galaxy:
One ranging from 
3500--6500~\AA\ (referred to as the blue setup hereafter) and one from
5000--7000~\AA\ (called the red setup hereafter), and to combine them after data reduction.

The volume phase holographic (VPH) grating GR900 was used to cover the
blue setup with a final reciprocal dispersion of $\sim$0.97~\AA\
pixel$^{-1}$ and a spectral resolution resulting in a FWHM of
5--6~\AA\ ($R\sim1000$).  In order to cover the red setup, we used
the VPH grating GR1300 with
a final reciprocal dispersion of $\sim$0.64~\AA\ pixel$^{-1}$ and a
spectral resolution with a FWHM of 3--4~\AA\ (R$\sim$1600).  
All observations were carried out between June 2012 and April 2013.

SALT is a telescope where during an observation the mirror remains 
at a fixed altitude and azimuth and the image of an astronomical target 
produced by the telescope is followed by the ``tracker'', which is located 
at the position of the prime focus (similar in operation to the 
Arecibo Radio Telescope). 
This results in only a limited observing window per target.
For this reason each observation
consisted usually of two exposures of about 1000~s each to fit a single
SALT visibility track. For the same reason the blue and red setups were
usually observed in different nights.

Spectra of Ar comparison arcs and a set of quartz tungsten halogen
flats were obtained immediately after each observation to calibrate
the wavelength scale and to correct for pixel-to-pixel variations.  A
set of spectrophotometric standard stars was observed during twilight
time for the relative flux calibration.  Since SALT has a variable
pupil size, an absolute flux calibration is not possible even using
spectrophotometric standard stars.  All details of the observations
are summarized in Table~\ref{tab:Obs}.  In total, 34 H\,{\sc ii}
regions in six galaxies were observed.

The primary reduction of the SALT data was done with the SALT science
pipeline \citep{Crawf10}.  After that, the bias- and gain-corrected and
mosaiced MOS data were reduced in the way described below.

\begin{table*}
\caption{Journal of the observations.} \label{tab:Obs}
\begin{tabular}{lcccccc} \hline
Galaxy & \MC{1}{c}{Date for}  & \MC{1}{c}{Date for} & \MC{1}{c}{Exposure time}  & \MC{1}{c}{Exposure time} & \MC{1}{c}{Seeing for} & \MC{1}{c}{Seeing for} \\
       & \MC{1}{c}{blue setup}& \MC{1}{c}{red setup}& \MC{1}{c}{for blue setup} & \MC{1}{c}{for red setup} & \MC{1}{c}{blue setup} & \MC{1}{c}{red setup} \\
 \hline
NGC\,1087  & ...        & 2012.12.20 &  ...                       & 2$\times$1000 &   ...          &  1.4$\arcsec$  \\
NGC\,2967  & 2013.02.02 & 2013.02.02 &  2$\times$1000             & 2$\times$1000 &   1.5$\arcsec$ &  1.2$\arcsec$  \\
NGC\,3023  & 2013.01.05 & 2013.01.06 &  2$\times$1000             & 2$\times$1000 &   2.5$\arcsec$ &  2.3$\arcsec$  \\
NGC\,4030  & 2013.03.20 & ...        &  2$\times$1000             & ...           &   1.0$\arcsec$ &  ...           \\
NGC\,4123  & 2013.04.29 & 2013.03.03 &  2$\times$1000             & 2$\times$910  &   1.7$\arcsec$ &  2.2$\arcsec$  \\
NGC\,4517A & 2012.06.16 & ...        &  1$\times$900+2$\times$725 & ...           &   1.7$\arcsec$ &  ...           \\
 \hline
\end{tabular}
\end{table*}

\subsection{Data reduction and line flux measurements}

Cosmic ray rejection was done using the IRAF\footnote{IRAF is
distributed by the National Optical Astronomical Observatories, which
are operated by the Association of Universities for Research in
Astronomy, Inc., under cooperative agreement with the National Science
Foundation.} task {\tt lacose$\_$spec} \citep{Dokkum2001}.  The
wavelength calibration was accomplished using the IRAF tasks {\tt
identify}, {\tt reidentify}, {\tt fitcoord}, and {\tt transform}. The
spectral data were divided by the illumination-corrected flat field in
order to correct for pixel-to-pixel sensitivity variations of the
detector.  After that two-dimensional spectra were extracted from the
MOS images for each slit.  The background subtraction was done using
the IRAF task {\tt background}.  Since we used a multi-slit mask, the
slits for the individual objects have a length of $\sim 5 - 20$~arcsec
and the background was fitted by a low-order polynomial function along
the spatial slit coordinate at each wavelength.  This allows us to
extract the flux from the H\,{\sc ii} regions only, without galactic
stellar background.  The spectra were corrected for sensitivity
effects using the Sutherland extinction curve and a sensitivity curve
obtained from observed standard star spectra.  Finally, all
two-dimensional spectra of a slit position obtained with the same
observational setup were averaged.

From each two-dimensional spectrum of the blue and red setups, one-dimensional 
spectra were extracted in spatial direction for each pixel along the slit. 
We refer to these
spectra as ``one-pixel-wide''.
The line fluxes 
([O\,{\sc ii}]$\lambda$$\lambda$3727,3729, 
[O\,{\sc iii}]$\lambda$4363,
 H${\beta}$,
[O\,{\sc iii}]$\lambda$4959, 
[O\,{\sc iii}]$\lambda$5007, 
[N\,{\sc ii}]$\lambda$6548,
H${\alpha}$, 
[N\,{\sc ii}]$\lambda$6584,
 [S\,{\sc ii}]$\lambda$6717, and 
[S\,{\sc ii}]$\lambda$6731) 
were then measured with IRAF or/and by fitting the lines with
Gaussians following \citet{PilyuginThuan2007} and
\citet{Pilyugin2010b}. 

\begin{figure}
\resizebox{1.00\hsize}{!}{\includegraphics[angle=000]{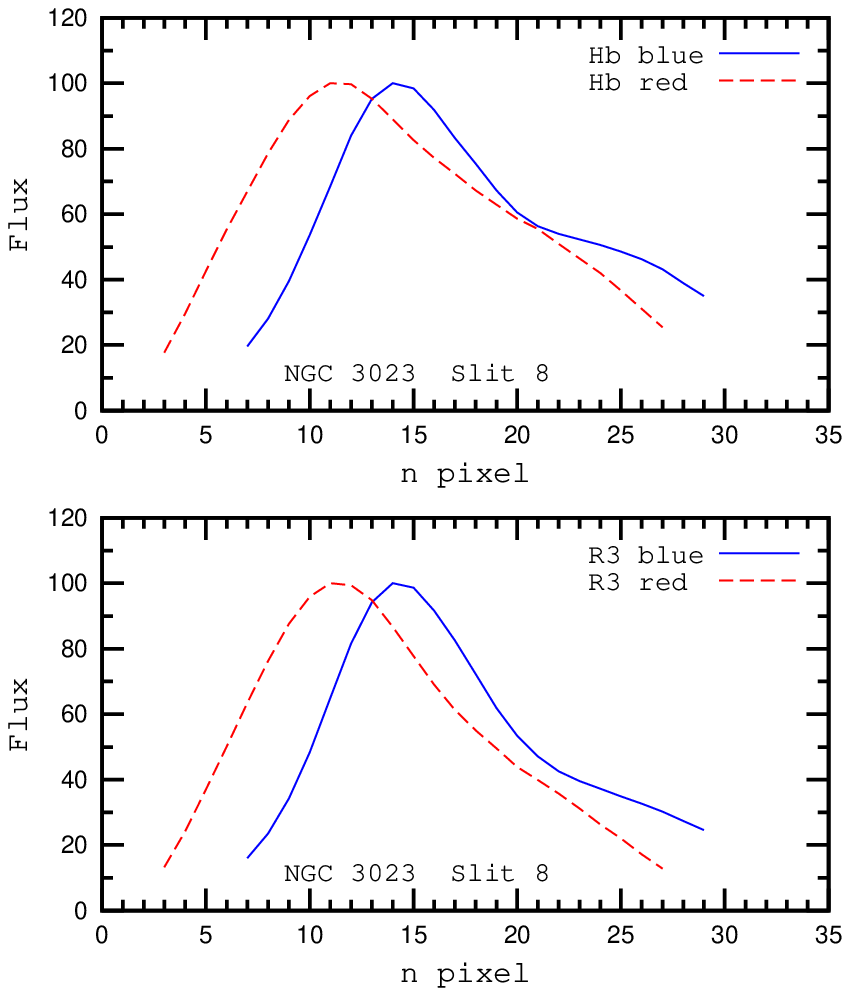}}
\caption{
The fluxes in the H$\beta$ (upper panel) and $R_3$ (lower panel) 
emission lines in the blue (solid lines) and red (dashed lines) spectra 
as a function of the pixel number along the slit for slit 8 in NGC~3023. 
The fluxes are in arbitrary units. 
}
\label{figure:j-flux-08}
\end{figure}

We first consider the distribution of the emission-line fluxes along
the slit.  The measured fluxes in the H$\beta$ and $R_3$ emission
lines in the blue and red one-pixel-wide spectra  for slit 8 in
NGC~3023 as a function of the pixel number along the slit are shown in
Fig.~\ref{figure:j-flux-08}.   Examination of
Fig.~\ref{figure:j-flux-08} shows that the position of the peak in the
H$\beta$ (and $R_3$) emission line in the red spectrum is shifted as
compared  to that in the blue spectrum by approximately three pixels.
This shows that the position of the slit of the red spectrum does not
coincide with the position of the slit for the blue spectrum. Instead
they are shifted in the direction along the slit by approximately
three pixels with respect to each other.  Inspection of
Fig.~\ref{figure:j-flux-08} also shows that the form of the
distribution of the H$\beta$ (and $R_3$) emission-line flux per pixel
in the red spectrum differs from that in the blue spectrum.  This
demonstrates that the position of the slit for the red spectrum is
also shifted in the direction perpendicular to the slit or that the
position angles of the red and blue spectrum are different.  This
prevents us from considering the lines of the blue and red spectra
together.  Therefore we derived the abundances using the lines from
the blue (or red) spectrum individually.

The full set of lines [O\,{\sc ii}]$\lambda$$\lambda$3727,3729,
H$\beta$, [O\,{\sc iii}]$\lambda$5007, H$\alpha$, and [N\,{\sc
ii}]$\lambda$6548 or H$\beta$, [O\,{\sc iii}]$\lambda$5007, H$\alpha$,
[N\,{\sc ii}]$\lambda$6584 and [S\,{\sc
ii}]$\lambda$$\lambda$6717,6731 is needed to correct for interstellar
reddening and to determine the oxygen abundances. For this reason only
slits that provide at least one of these line sets in either the blue
or red setup are chosen for further study.
As the actual spectral coverage for each of our slits varies 
slightly depending on the position of a given slit on the mask, in some
cases the H$\beta$+[O\,{\sc iii}]$\lambda$5007 lines of the red spectra and 
the H$\alpha$+[N\,{\sc ii}]$\lambda$6548 lines of the blue spectra 
are shifted beyond the actual spectral coverage of the slit. This is
the reason why the number of the presented observations of H\,{\sc ii} regions 
varies from one in NGC~4030 to up to 9 in NGC~1087.

We constructed the aperture for the blue and red spectra by averaging
the seven one-pixel-wide spectra near the flux maximum.  
As mentioned before, the emission from the underlying stellar population 
of the galactic disk was subtracted during the background correction, 
i.e., we removed the stellar continuum averaged along the spatial 
slit coordinates near the H\,{\sc ii} region.
Since the continuum in the spectra of our H\,{\sc ii} regions is sufficiently
weak or undetectable, we neglected possible stellar absorption by the stellar 
populations of the H\,{\sc ii} regions.
The measured emission fluxes $F$ were corrected for interstellar reddening. 
We obtained the extinction coefficient C(H$\beta$) using the theoretical
H$\alpha$-to-H$\beta$ ratio (= 2.878) and the analytical approximation
to the Whitford interstellar reddening law of \citet{Izotov1994}.


\begin{table*}
{\tiny
\caption[]{\label{table:linesblue}
The dereddened emission line fluxes (in units of the H$\beta$ line flux) 
and the extinction coefficient C(H$\beta$) in the blue spectra of a 
sample of the target H\,{\sc ii} regions in NGC~3023. 
}
\begin{center}
\begin{tabular}{cccccccccc} \hline \hline
Slit                                         & 
R.A.$^{a}$                                    &
DEC.$^{a}$                                    &
[O\,{\sc ii}]$\lambda$3727,3729              & 
[O\,{\sc iii}]$\lambda$4363                  & 
[O\,{\sc iii}]$\lambda$5007                  & 
[N\,{\sc ii}]$\lambda$6584                   & 
[S\,{\sc ii}]$\lambda$6717                   & 
[S\,{\sc ii}]$\lambda$6731                   & 
C(H$\beta$)                                  \\  \hline
\multicolumn{10}{c}{NGC~2967} \\
\hline
 17  &  145.503249 &    0.340438 &   2.712 $\pm$ 0.164 &                     &   0.371 $\pm$ 0.033 &   0.772 $\pm$ 0.044 &                     &                     &   0.579  \\ 
 33  &  145.516052 &    0.316465 &   4.603 $\pm$ 0.662 &                     &   1.788 $\pm$ 0.180 &   0.621 $\pm$ 0.041 &                     &                     &   0.704  \\ 
\hline
\multicolumn{10}{c}{NGC~3023} \\
\hline
  8  &  147.476616 &    0.616676 &   1.627 $\pm$ 0.051 &   0.072 $\pm$ 0.003 &   6.117 $\pm$ 0.200 &                     &                     &                     &   0.297  \\ 
 12  &  147.474086 &    0.615905 &   2.600 $\pm$ 0.084 &   0.030 $\pm$ 0.003 &   3.280 $\pm$ 0.109 &                     &                     &                     &   0.317  \\ 
 14  &  147.468182 &    0.619303 &   3.662 $\pm$ 0.120 &                     &   1.654 $\pm$ 0.057 &   0.474 $\pm$ 0.021 &                     &                     &   0.345  \\ 
\hline
\multicolumn{10}{c}{NGC~4030} \\
\hline
 46  &  180.111544 &   -1.074890 &   2.340 $\pm$ 0.104 &                     &   0.820 $\pm$ 0.034 &   0.849 $\pm$ 0.038 &                     &                     &   0.524  \\ 
\hline
\multicolumn{10}{c}{NGC~4123} \\
\hline
 14  &  182.032991 &    2.880344 &   2.281 $\pm$ 0.131 &                     &   0.457 $\pm$ 0.031 &   0.858 $\pm$ 0.036 &                     &                     &   0.535  \\ 
 20  &  182.030381 &    2.882922 &   2.324 $\pm$ 0.139 &                     &   0.896 $\pm$ 0.045 &   0.817 $\pm$ 0.039 &                     &                     &   0.403  \\ 
 21  &  182.031526 &    2.888546 &   2.322 $\pm$ 0.146 &                     &   0.473 $\pm$ 0.036 &   0.903 $\pm$ 0.040 &                     &                     &   0.689  \\ 
 22  &  182.024433 &    2.888972 &   2.925 $\pm$ 0.166 &                     &   0.781 $\pm$ 0.051 &   0.694 $\pm$ 0.032 &   0.483 $\pm$ 0.026 &   0.331 $\pm$ 0.017 &   0.485  \\ 
 26  &  182.022560 &    2.908112 &   1.771 $\pm$ 0.088 &                     &   3.457 $\pm$ 0.142 &   0.217 $\pm$ 0.014 &                     &                     &   0.113  \\ 
\hline
\multicolumn{10}{c}{NGC~4517A} \\
\hline
 44  &  188.108016 &    0.381373 &   4.566 $\pm$ 0.224 &                     &   0.936 $\pm$ 0.050 &   0.269 $\pm$ 0.020 &                     &                     &   0.402  \\ 
 56  &  188.116422 &    0.390758 &   2.310 $\pm$ 0.084 &                     &   2.109 $\pm$ 0.096 &   0.305 $\pm$ 0.021 &                     &                     &   0.194  \\ 
\hline 
\end{tabular}\\
\end{center}
\begin{flushleft}
$^a$ in degrees (J2000).\\
\end{flushleft}
}
\end{table*}

\begin{table*}
\caption[]{\label{table:linesred}
The dereddened emission line fluxes (in units of the H$\beta$ line flux) 
and the extinction coefficient C(H$\beta$) in the red spectra of a 
sample of the target H\,{\sc ii} regions in our galaxy sample. 
}
\begin{center}
\begin{tabular}{cccccccc} \hline \hline
Slit                                         & 
R.A.$^{a}$                                   &
DEC.$^{a}$                                   &
[O\,{\sc iii}]$\lambda$5007                  & 
[N\,{\sc ii}]$\lambda$6584                   & 
[S\,{\sc ii}]$\lambda$6717                   & 
[S\,{\sc ii}]$\lambda$6731                   & 
C(H$\beta$)                                  \\  \hline
\multicolumn{8}{c}{NGC~1087} \\
\hline
 15  &   41.605348 &   -0.482317 &   0.794 $\pm$  0.034 &   0.633 $\pm$  0.030 &   0.410 $\pm$  0.018 &   0.292 $\pm$  0.013 &   0.353  \\ 
 16  &   41.609253 &   -0.490142 &   0.494 $\pm$  0.020 &   0.750 $\pm$  0.029 &   0.479 $\pm$  0.017 &   0.336 $\pm$  0.013 &   0.282  \\ 
 17  &   41.607925 &   -0.493503 &   0.400 $\pm$  0.031 &   0.664 $\pm$  0.026 &   0.377 $\pm$  0.018 &   0.259 $\pm$  0.014 &   0.306  \\ 
 23  &   41.604852 &   -0.497109 &   0.211 $\pm$  0.011 &   0.935 $\pm$  0.043 &   0.373 $\pm$  0.017 &   0.278 $\pm$  0.012 &   0.380  \\ 
 28  &   41.610902 &   -0.500035 &   0.429 $\pm$  0.033 &   0.754 $\pm$  0.026 &   0.443 $\pm$  0.020 &   0.334 $\pm$  0.015 &   0.252  \\ 
 30  &   41.610902 &   -0.486017 &   0.427 $\pm$  0.017 &   0.696 $\pm$  0.033 &   0.320 $\pm$  0.016 &   0.225 $\pm$  0.010 &   0.340  \\ 
 31  &   41.603092 &   -0.504827 &   0.449 $\pm$  0.014 &   0.764 $\pm$  0.027 &   0.275 $\pm$  0.009 &   0.206 $\pm$  0.007 &   0.443  \\ 
 32  &   41.604192 &   -0.512645 &   0.932 $\pm$  0.036 &   0.615 $\pm$  0.031 &   0.464 $\pm$  0.027 &   0.322 $\pm$  0.023 &   0.417  \\ 
 36  &   41.609197 &   -0.515824 &   0.406 $\pm$  0.021 &   0.589 $\pm$  0.025 &   0.402 $\pm$  0.016 &   0.280 $\pm$  0.016 &   0.251  \\ 
\hline
\multicolumn{8}{c}{NGC~2967} \\
\hline
 14  &  145.509118 &    0.341424 &   0.129 $\pm$  0.013 &   0.840 $\pm$  0.031 &   0.309 $\pm$  0.013 &   0.226 $\pm$  0.009 &   0.900  \\ 
 23  &  145.522704 &    0.332752 &   0.195 $\pm$  0.060 &   0.820 $\pm$  0.046 &   0.395 $\pm$  0.024 &   0.270 $\pm$  0.019 &   1.040  \\ 
 28  &  145.521731 &    0.326422 &   0.930 $\pm$  0.048 &   0.613 $\pm$  0.045 &   0.328 $\pm$  0.031 &   0.216 $\pm$  0.019 &   0.970  \\ 
 29  &  145.520986 &    0.328543 &   0.294 $\pm$  0.070 &   0.676 $\pm$  0.031 &   0.350 $\pm$  0.029 &   0.252 $\pm$  0.035 &   0.734  \\ 
 37  &  145.502932 &    0.349615 &   0.738 $\pm$  0.126 &   0.689 $\pm$  0.061 &   0.608 $\pm$  0.049 &   0.295 $\pm$  0.050 &   0.902  \\ 
 39  &  145.505741 &    0.355989 &   1.596 $\pm$  0.199 &   0.530 $\pm$  0.059 &   0.451 $\pm$  0.064 &   0.229 $\pm$  0.055 &   0.644  \\ 
\hline
\multicolumn{8}{c}{NGC~3023} \\
\hline
  8  &  147.476616 &    0.616676 &   5.680 $\pm$  0.204 &   0.092 $\pm$  0.004 &   0.122 $\pm$  0.005 &   0.089 $\pm$  0.004 &   0.255  \\ 
 12  &  147.474086 &    0.615905 &   3.118 $\pm$  0.103 &   0.221 $\pm$  0.008 &   0.225 $\pm$  0.009 &   0.165 $\pm$  0.007 &   0.181  \\ 
 14  &  147.468182 &    0.619303 &   1.740 $\pm$  0.064 &   0.414 $\pm$  0.017 &   0.456 $\pm$  0.019 &   0.332 $\pm$  0.016 &   0.222  \\ 
 18  &  147.480337 &    0.619130 &   3.482 $\pm$  0.130 &   0.147 $\pm$  0.006 &   0.225 $\pm$  0.010 &   0.146 $\pm$  0.006 &   0.058  \\ 
 19  &  147.463502 &    0.612667 &   2.479 $\pm$  0.110 &   0.429 $\pm$  0.041 &   0.500 $\pm$  0.044 &   0.434 $\pm$  0.044 &   0.541  \\ 
 21  &  147.457781 &    0.612337 &   1.872 $\pm$  0.071 &   0.309 $\pm$  0.016 &   0.484 $\pm$  0.021 &   0.337 $\pm$  0.016 &   0.104  \\ 
 22  &  147.486919 &    0.623162 &   2.748 $\pm$  0.094 &   0.136 $\pm$  0.009 &   0.240 $\pm$  0.011 &   0.167 $\pm$  0.009 &   0.078  \\ 
\hline
\multicolumn{8}{c}{NGC~4123} \\
\hline
  6  &  182.046371 &    2.878087 &   0.267 $\pm$  0.012 &   1.523 $\pm$  0.051 &   0.334 $\pm$  0.011 &   0.363 $\pm$  0.012 &   1.086  \\ 
 18  &  182.061021 &    2.864815 &   0.455 $\pm$  0.043 &   0.816 $\pm$  0.039 &   0.584 $\pm$  0.033 &   0.478 $\pm$  0.035 &   0.424  \\ 
\hline 
\end{tabular}\\
\end{center}
\begin{flushleft}
$^a$ in degrees (J2000). \\ 
\end{flushleft}
\end{table*}

The dereddened emission-line fluxes in the averaged spectra of the
target H\,{\sc ii} regions are listed in Table \ref{table:linesblue}
for the blue spectra and in Table \ref{table:linesred} for the red
spectra.  The theoretical ratio of [N\,{\sc ii}]$\lambda$6584/[N\,{\sc
ii}]$\lambda$6548 is constant and close to 3 \citep{Storey2000} since
those lines originate from transitions from the same energy level.
Since the [N\,{\sc ii}]$\lambda$6584 line measurements are more
reliable than the [N\,{\sc ii}]$\lambda$6548 line measurements the
value of $N_2$ is estimated as $N_2$  = 1.33$\times$[N\,{\sc
ii}]$\lambda$6584 unless indicated otherwise.  Similarly, the value of
$R_3$ can be estimated as $R_3$  = 1.33$\times$[O\,{\sc
iii}]$\lambda$5007 since the [O\,{\sc iii}]$\lambda$5007 and [O\,{\sc
iii}]$\lambda$4959 lines originate also from transitions from the same
energy level and their flux ratio is very close to 3
\citep{Storey2000,SHOC}.  Therefore, the [N\,{\sc ii}]$\lambda$6548
and $\lambda$4959 lines are not included in Table
\ref{table:linesblue} and Table \ref{table:linesred}.

The uncertainty of the emission-line flux $\varepsilon_{line}$ is
estimated taking into account the uncertainty of the continuum level,
errors in the line flux and the uncertainty in the sensitivity curve
\citep[see][for details]{SHOC}.  The uncertainty of the continuum,
$\varepsilon_{cont}$, is determined in the region near the emission
line where the continuum is approximated by a linear fit.  The
line-flux uncertainty, $\varepsilon_{flux}$, is estimated as the
deviation from a Gaussian profile.  The uncertainty in the sensitivity
curve, $\varepsilon_{sc}$, is less than 2 -- 3\% in all considered
wavelength ranges \citep[see, e.g.,][]{Kniazev12}. We adopt the
maximum value of the relative uncertainty $\varepsilon_{sc}$ = 0.03.

\section{Photometry}

To estimate the deprojected galactocentric distance (normalized to the
optical isophotal radius $R_{25}$) of the H\,{\sc ii} region from its
coordinates on the celestial sphere one needs to know the values of
the inclination, $i$, the position angle of the major axis, {\it PA},
and the isophotal radius of a galaxy, $R_{25}$.  It is common practice
to take those values from \citet[][thereafter RC3]{RC3} or from the
{\sc leda} database. However, some values from those sources show a
significant difference for galaxies from our list. For example, the
isophotal radius of NGC~4517A in the RC3 is larger by a factor of
two than that given in the {\sc leda} database. Therefore we obtained
our own estimates of the values of $i$, {\it PA}, and  $R_{25}$ for
our target galaxies.

\begin{figure}
\resizebox{1.00\hsize}{!}{\includegraphics[angle=000]{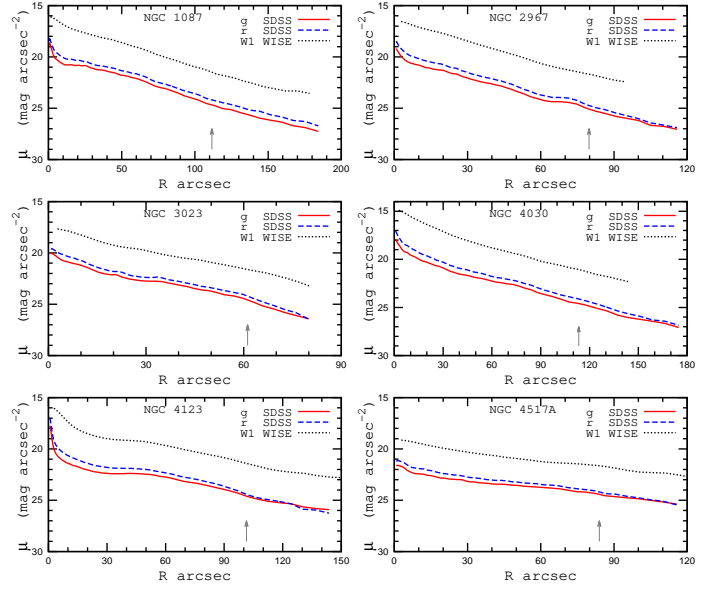}}
\caption{
The observed surface-brightness profiles of our galaxies in the $g$
and $r$ bands of the {\it SDSS} photometric system and in the $W1$
band of the {\it WISE} photometric system. The X-axis shows the
galactocentric radius in arcsec, and the Y-axis the surface brightness
in mag arcsec$^{-2}$.  The optical isophotal radius $R_{25}$ is marked
by an arrow. 
}
\label{figure:profilesall}
\end{figure}

\begin{figure}
\resizebox{1.00\hsize}{!}{\includegraphics[angle=000]{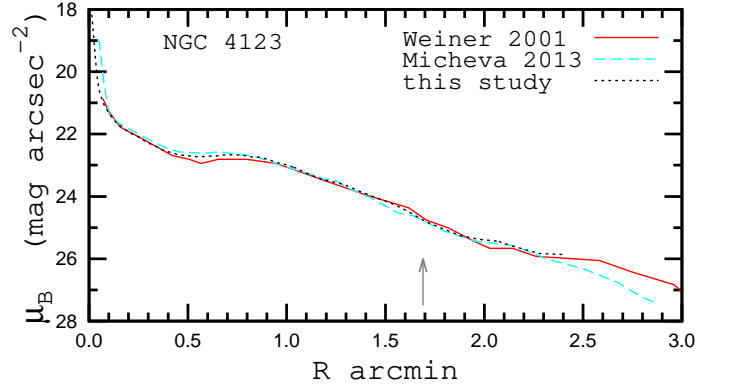}}
\caption{
Comparison between the measured surface-brightness profiles 
of NGC~4123 in the $B$ band reported by \citet{Weiner2001} (solid 
line), by \citet{Micheva2013} (long-dashed line), and obtained here 
(short-dashed line).  The X-axis shows the galactocentric radius in
arcmin, and the Y-axis the surface brightness in mag arcsec$^{-2}$. 
The arrow indicates the optical isophotal radius $R_{25}$. 
}
\label{figure:NGC4123Weiner}
\end{figure}

\begin{figure}
\resizebox{1.00\hsize}{!}{\includegraphics[angle=000]{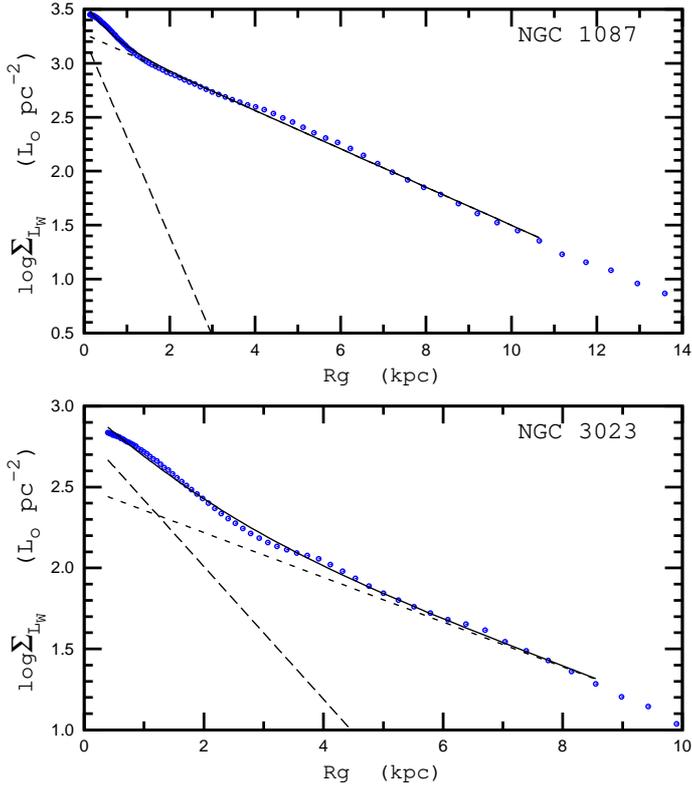}}
\caption{
The patterns resulting from the bulge-disk decomposition of our target
galaxies (X-axis: galactocentric radius in kpc, Y-axis: logarithm of
the central surface brightness for a face-on galaxy orientation in 
solar luminosities per pc$^2$).
Each panel shows the decomposition assuming a purely exponential
profile for the disk.  The measured surface profile is plotted using
gray (blue) circles.  The bulge contribution is shown by a dotted
line, the disk contribution by a dashed line, and the total (bulge +
disk) fit by a solid line.  
\label{figure:decomposit}
}
\end{figure}

We analyzed the publicly available photometric maps in the infrared
$W1$ band (with an isophotal wavelength of 3.4 $\mu$m) obtained by the {\it
Wide-field Infrared Survey Explorer (WISE)} \citep{Wright2010} and in
the $g$ and $r$ bands obtained by the {\it Sloan Digital Sky Survey}
\citep[{\it SDSS}; data release 9 (DR9),][]{Ahn2012}.  We derived the
surface-brightness profile and disk orientation parameters in three
photometric bands for each galaxy.  The determinations of the
surface-brightness profile, position angle, and ellipticity were
performed for each band separately in the way described in
\citet{Pilyugin2014b}.  But for NGC~3023 we were not able to estimate
reliable values of the position angle and ellipticity for the $g$ and
$r$ bands. Therefore, for this galaxy the values of the position angle
and ellipticity obtained for the $W1$ band were used for the
construction of the surface-brightness profiles in all three filters.

It should be noted that the {\it WISE} and {\it SDSS} surveys are
sufficiently deep for our surface-brightness profiles to extend beyond
the optical isophotal radii $R_{25}$.  The obtained surface-brightness
profiles are shown in Fig.~\ref{figure:profilesall}.   The adopted
inclinations and position angles are given in Table
\ref{table:sample}. 

The value of the isophotal radius is derived from the obtained
surface-brightness profiles in the $g$ and $r$ bands.
Surface-brightness measurements were corrected for foreground Galactic
extinction using the $A_V$ values from the recalibration by
\citet{Schlafly2011} of the extinction maps of \citet{Schlegel1998}
and the extinction curve of \citet{Cardelli1989}, assuming a ratio of
total to selective extinction of $R_{V} = A_{V}/E_{B-V} = 3.1$. The
$A_V$ values given in the NASA Extragalactic Database {\sc ned} were
used.  Afterwards the surface-brightness measurements were corrected
for the inclination.  The measurements in the {\it SDSS} filters $g$
and $r$ were converted to $B$-band magnitudes, and the $AB$ magnitudes
were reduced to the Vega photometric system using the conversion
relations and solar magnitudes of \citet{Blanton2007}.  First, the
$B$-band magnitudes were obtained from the $g$ and $r$ magnitudes
\begin{equation} B_{AB} = g + 0.2354 + 0.3915\;[(g-r)-0.6102] ,
\label{equation:bgr} \end{equation} where the $B_{AB}$, $g$, and $r$
magnitudes in Eq.~(\ref{equation:bgr}) are in the $AB$ photometric
system.  Then, the $AB$ magnitudes were reduced to the Vega
photometric system \begin{equation} B_{Vega} = B_{AB} + 0.09  .
\label{equation:VegaAB} \end{equation} The obtained isophotal radii
are given in Table \ref{table:sample}. 

Surface brightness profiles of the galaxy NGC~4123 in the $B$ band
were published by \citet{Weiner2001} and \citet{Micheva2013}.
Fig.~\ref{figure:NGC4123Weiner} shows the comparison between their
profiles with the one derived here from the photometric imaging
data in the {\it SDSS} $g$ and $r$ bands. 

We performed a bulge-disk decomposition of the observed
surface-brightness profiles using a purely exponential disk (PED)
approximation in the same way as in \citet{Pilyugin2014b}.
Exponential profiles were used to fit the observed disk
surface-brightness profiles, and the bulge profiles were fitted with a
general S\'{e}rsic profile.  The observed surface-brightness profiles
of five galaxies from our sample are fitted satisfactorily well.
The upper panel of Figure~\ref{figure:decomposit} shows the bulge-disk
decomposition of the galaxy NGC~1087 as an example.  The measured
surface profile is marked by circles.  The fit to the bulge
contribution is shown by a dotted line, the fit to the disk by a
dashed line, and the total (bulge + disk) fitting by a solid line.
Table \ref{table:sample} lists the parameters of the disk
surface-brightness profiles of those galaxies in the $W1$ band: the
logarithm of the central surface brightness of the disk in the $W1$
band reduced to a face-on galaxy orientation in terms of $L_{\sun}$
pc$^{-2}$ and the disk scale length in the $W1$ band, $h_{W1}$ in kpc.
Those values are parameters of the exponential disk
approximation, described in detail in \citet{Pilyugin2014b}. 

For the galaxy NGC~3023, we could not determine a reliable disk scale
length, $h_{W1}$, and central surface brightness of the disk,
$(\Sigma_{L_{W1}})_{0}$.  The lower panel of
Figure~\ref{figure:decomposit} shows the surface-brightness-profile
fit for this galaxy. The disk contribution to the surface
brightness is close to the observed surface-brightness profile over a
small interval of radial distances only (in fact, this is a
bulge-dominated galaxy). Therefore, the values of the disk scale
length and central surface brightness of the disk are questionable.

\section{Abundances}

\subsection{Abundance determination}
\label{txt:abund}

\begin{figure}
\resizebox{1.00\hsize}{!}{\includegraphics[angle=000]{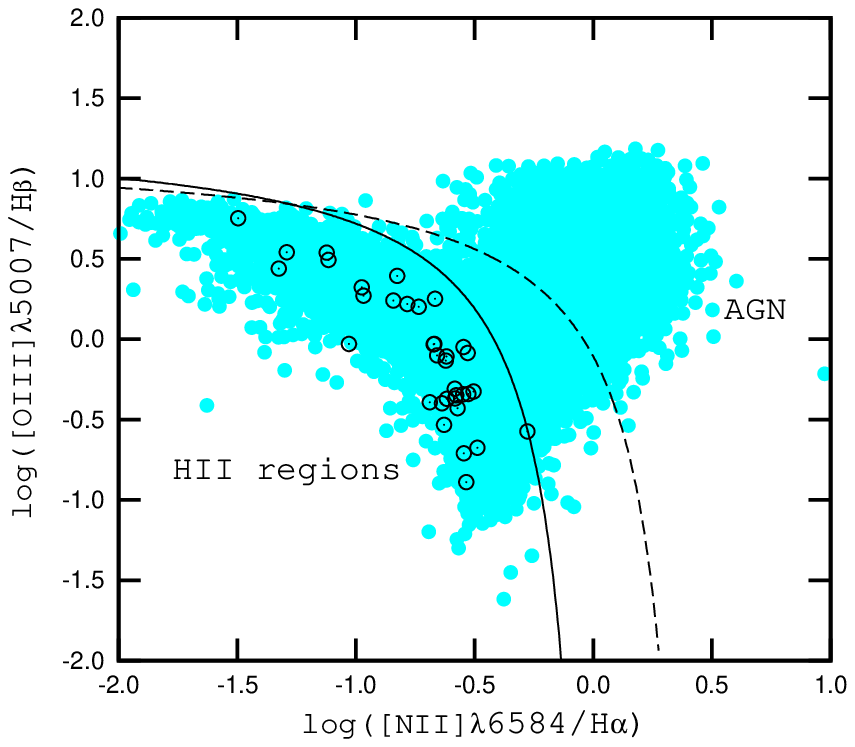}}
\caption{
The [N\,{\sc ii}]$\lambda$6584/H$\alpha$ versus 
[O\,{\sc iii}]$\lambda$5007/H$\beta$ diagram.  The symbols denote 
results for the measured H\,{\sc ii} regions in our target galaxies. 
The solid line separates objects with H\,{\sc ii} region spectra 
from those containing an AGN according to \citet{Kauffmann2003}, 
while the dashed line represents the same separation according to 
the work by \citet{Kewley2001}. 
The gray (light-blue) filled circles show a large sample of 
emission-line {\it SDSS} galaxies from \citet{Thuan2010}. 
}
\label{figure:bpt}
\end{figure}

\citet{Baldwin1981} proposed the [OIII]$\lambda$5007/H${\beta}$ vs.\
[NII]$\lambda$6584/H${\alpha}$ diagram (the so-called BPT
classification diagram) which is often used to distinguish between
star-forming regions and AGNs. The exact location of the dividing line
between star-forming regions and AGNs is still controversial
\citep[see, e.g.,][]{Kewley2001,Kauffmann2003}.  Fig.~\ref{figure:bpt}
shows the positions of our targets (open circles) in the BPT
classification diagram. The solid line is the dividing line between
star-forming regions and AGNs according to \citet{Kauffmann2003},
while the dashed line is the same line according to
\citet{Kewley2001}.  Regardless of which line is adopted,
Fig.~\ref{figure:bpt} shows that all our objects are H\,{\sc ii}
regions and their oxygen abundances can be estimated using standard
techniques. 

The $T_{e}$-based oxygen (O/H)$_{T_{e}}$ abundances of the H\,{\sc ii}
regions with the detected auroral line [O\,{\sc iii}]$\lambda$4363
were determined using the equations for the $T_{e}$-method from
\citet{Pilyugin2010a,Pilyugin2012}. 

A new method (called the ``$C$ method'') for oxygen and nitrogen
abundance determinations from strong emission lines was recently
suggested \citep{Pilyugin2012,Pilyugin2014a}.  In our red spectra,
we measured the strong lines $R_3$, $N_2$, and $S_2$, which allowed
us to determine the oxygen (O/H)$_{C_{\rm NS}}$ abundances using those strong
lines.  In some of our blue spectra, the strong lines $R_2$, $R_3$, and 
$N_2$ were measured and applied to determine the oxygen (O/H)$_{C_{\rm ON}}$
abundances.

\subsection{The robustness and precision of the abundance determination}

\begin{figure}
\resizebox{1.00\hsize}{!}{\includegraphics[angle=000]{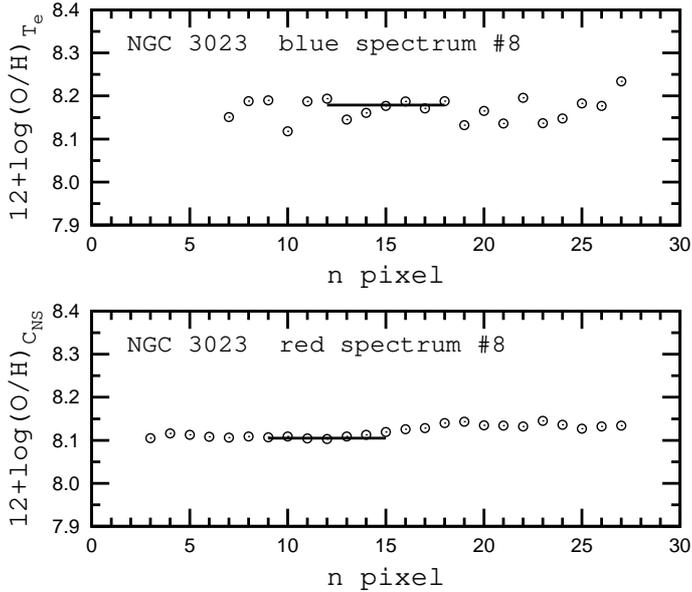}}
\caption{
The oxygen abundances as a function of the number of the pixel along the slit 
for the slit 8 in the NGC~3023. 
The circles in the upper panel show the oxygen abundances determined from the 
individual blue one-pixel-width spectra through the $T_{e}$ method. The solid 
line is the abundance obtained from the integrated seven-pixel-width spectrum 
through the $T_{e}$ method.
The circles in the lower panel show the oxygen abundances determined from the 
individual red one-pixel-width spectra through the $C_{NS}$ method. The solid 
line is the abundance obtained from the integrated seven-pixel-width spectrum 
through the $C_{NS}$ method.
}
\label{figure:j-oh-08}
\end{figure}

The emission-line fluxes measured in the one-pixel-wide spectra represent the radiation of a 
small part of the  H\,{\sc ii} region. One would expect that the $T_{e}$-based abundances  
in a given H\,{\sc ii} region derived from the spectra of different areas 
on the  H\,{\sc ii} region image should be the same or at least should 
be close to each other. Is this the case for the abundances estimated from the counterpart 
method? To clarify this matter we have estimated the oxygen abundances from the individual 
one-pixel-wide spectra and considered the variations in those abundances. 

Fig.~\ref{figure:j-oh-08} shows the distribution of the oxygen abundances along the slit for 
the bright, extend H\,{\sc ii} region (\# 8) in NGC~3023, i.e., the abundance estimated from 
the individual one-pixel-wide spectra as a function of the number of the pixel along the slit. 
The auroral line  $R$  =  [O\,{\sc iii}]$\lambda$4363 is detected in around 20 individual 
one-pixel-wide spectra of this H\,{\sc ii} region. The circles in the upper panel of the 
Fig.~\ref{figure:j-oh-08}  show the oxygen abundances determined from the individual blue 
one-pixel-wide spectra through the $T_{e}$ method. The solid line is the $T_{e}$-based abundance 
obtained from the seven-pixel-wide spectrum. The circles in the lower panel  of the 
Fig.~\ref{figure:j-oh-08} show the oxygen abundances determined from the individual red 
one-pixel-wide spectra through the $C_{NS}$ method. The solid line represents the abundance obtained 
from the seven-pixel-width spectrum through the $C_{NS}$ method.

Inspection of the upper panel of Fig.~\ref{figure:j-oh-08} confirms that the $T_{e}$-based abundances  
are independent from the position in the H\,{\sc ii} region image at which the measurement was taken. 
Examination of the lower panel of Fig.~\ref{figure:j-oh-08} shows that the abundances 
determined from the individual one-pixel-wide spectra through the $C$ method are close 
to each other and are close to the abundance obtained from the integrated seven-pixel-wide spectrum. 
A similar picture was found for other  H\,{\sc ii} regions. 
Thus, the  $C$-based abundances are robust and independent from the specific area covered by the measurement in an  H\,{\sc ii} region 
image, i.e., the $C$ method produces a reliable oxygen abundance even if a 
spectrum of only part of an H\,{\sc ii} region is used.  

The scatter in individual $C$-based abundances is even lower than the scatter in  individual $T_{e}$-based 
abundances, i.e., the formal uncertainty in the  $C$-based abundances is lower than that in 
the $T_{e}$-based abundances. This can be attributed to the fact that the measurements of the weak auroral lines 
used in the $T_{e}$ method can involve larger errors than the measurements of the strong lines used in the 
$C$ method. It should be noted, however, that the true uncertainty in the $C$-based oxygen abundance does not only depend 
on the accuracy of the strong-line measurements in the spectrum of the target  H\,{\sc ii} region 
but also on the reliability of the abundance determinations in the reference  H\,{\sc ii} regions. 
Our current sample of reference H\,{\sc ii} regions (our standard reference sample from 2013)  contains 250  H\,{\sc ii} regions 
for which the absolute differences in the oxygen abundances (O/H)$_{C_{\rm ON}}$  -- (O/H)$_{T_{e}}$ and 
(O/H)$_{C_{\rm NS}}$  -- (O/H)$_{T_{e}}$ and in the nitrogen abundances (N/H)$_{C_{\rm ON}}$  -- (N/H)$_{T_{e}}$ and
(N/H)$_{C_{\rm NS}}$  -- (N/H)$_{T_{e}}$ are less than 0.1~dex \citep{Pilyugin2014a}. 
Thus the true uncertainty in the $C$-based oxygen abundance may be up to around 0.1 dex even if the formal   
error due to the uncertainties in the strong-line measurement is small. Therefore we assume that 
the uncertainties in the obtained oxygen abundances in our investigated H\,{\sc ii} regions in the current paper can exceed 
0.1 dex although the formal error caused by the uncertainty in the line fluxes measurement is lower.

\begin{table}
\caption[]{\label{table:abundance}
Oxygen abundances in the H\,{\sc ii} regions in the disks of our sample 
of galaxies }
\begin{center}
\begin{tabular}{cccll} \hline \hline
Slit                & 
$R/R_{25}$           & 
12+log(O/H)$^a$         &
Method              & 
spectrum           \\  \hline
\multicolumn{5}{c}{NGC~1087} \\
\hline
  15  &  0.527  &  8.44  &  $C_{\rm NS}$       &  red  \\        
  16  &  0.359  &  8.48  &  $C_{\rm NS}$       &  red  \\        
  17  &  0.232  &  8.50  &  $C_{\rm NS}$       &  red  \\        
  23  &  0.050  &  8.61  &  $C_{\rm NS}$       &  red  \\        
  28  &  0.327  &  8.50  &  $C_{\rm NS}$       &  red  \\        
  30  &  0.506  &  8.52  &  $C_{\rm NS}$       &  red  \\        
  31  &  0.231  &  8.55  &  $C_{\rm NS}$       &  red  \\        
  32  &  0.444  &  8.42  &  $C_{\rm NS}$       &  red  \\        
  36  &  0.603  &  8.48  &  $C_{\rm NS}$       &  red  \\  
\hline
\multicolumn{5}{c}{NGC~2967} \\
\hline
  14  &  0.232  &  8.66  &  $C_{\rm NS}$       &  red  \\        
  17  &  0.554  &  8.53  &  $C_{\rm ON}$       &  blue \\        
  23  &  0.454  &  8.60  &  $C_{\rm NS}$       &  red  \\        
  28  &  0.575  &  8.45  &  $C_{\rm NS}$       &  red  \\        
  29  &  0.494  &  8.55  &  $C_{\rm NS}$       &  red  \\        
  33  &  0.936  &  8.35  &  $C_{\rm ON}$       &  blue \\        
  37  &  0.785  &  8.44  &  $C_{\rm NS}$       &  red  \\        
  39  &  0.949  &  8.36  &  $C_{\rm NS}$       &  red  \\        
\hline
\multicolumn{5}{c}{NGC~3023} \\
\hline
  08  &  0.449  &  8.16  &  $T_{e}$        &  blue  \\        
      &         &  8.10  &  $C_{\rm NS}$       &  red   \\        
  12  &  0.336  &  8.16  &  $T_{e}$        &  blue  \\        
      &         &  8.22  &  $C_{\rm NS}$       &  red   \\        
  14  &  0.110  &  8.34  &  $C_{\rm ON}$       &  blue  \\        
      &         &  8.30  &  $C_{\rm NS}$       &  red   \\        
  18  &  0.665  &  8.09  &  $C_{\rm NS}$       &  red   \\        
  19  &  0.575  &  8.22  &  $C_{\rm NS}$       &  red   \\        
  21  &  0.826  &  8.09  &  $C_{\rm NS}$       &  red   \\        
  22  &  1.117  &  7.99  &  $C_{\rm NS}$       &  red   \\        
\hline
\multicolumn{5}{c}{NGC~4030} \\
\hline
  46  &  0.910  &  8.52  &  $C_{\rm ON}$       &  blue  \\        
\hline
\multicolumn{5}{c}{NGC~4123} \\
\hline
   6  &  0.001  &  8.60  &  $C_{\rm NS}$       &  red   \\        
  14  &  0.528  &  8.54  &  $C_{\rm ON}$       &  blue  \\        
  18  &  0.711  &  8.47  &  $C_{\rm NS}$       &  red   \\        
  20  &  0.629  &  8.51  &  $C_{\rm ON}$       &  blue  \\        
  21  &  0.646  &  8.54  &  $C_{\rm ON}$       &  blue  \\        
  22  &  0.889  &  8.44  &  $C_{\rm NS}$       &  blue  \\        
      &         &  8.47  &  $C_{\rm ON}$       &  blue  \\        
  26  &  1.399  &  8.26  &  $C_{\rm ON}$       &  blue  \\        
\hline
\multicolumn{5}{c}{NGC~4517A} \\
\hline
  44  &  0.603  &  7.95  &  $C_{\rm ON}$       &  blue  \\        
  56  &  0.082  &  8.32  &  $C_{\rm ON}$       &  blue  \\        
\hline
\hline 
\MC{5}{l}{$^a$The standard error for the C-based methods is around 0.1~dex.}\\
\MC{5}{l}{See Section~\ref{txt:abund} for more details. }\\
\end{tabular}\\
\end{center}
\end{table}

\subsection{Radial gradients}

The radial distribution of the oxygen abundances across the disk
within the isophotal radius in each of our target galaxies was 
fitted with the following equation:
\begin{equation}
12+\log({\rm O/H})  = 12+\log({\rm O/H})_{R_{0}} + C_{\rm O/H} \times (R/R_{25}) ,
\label{equation:grado}
\end{equation} 
where 12 + log(O/H)$_{R_{0}}$ is the oxygen abundance at $R_{0}$ = 0,
i.e., the extrapolated central oxygen abundance, C$_{\rm O/H}$, is the
slope of the oxygen abundance gradient expressed in terms of
dex~$R_{\rm 25}^{-1}$, and $R$/$R_{\rm 25}$ is the fractional radius
(the galactocentric distance normalized to the disk's isophotal radius
$R_{25}$).

\begin{figure*}
\resizebox{1.00\hsize}{!}{\includegraphics[angle=000]{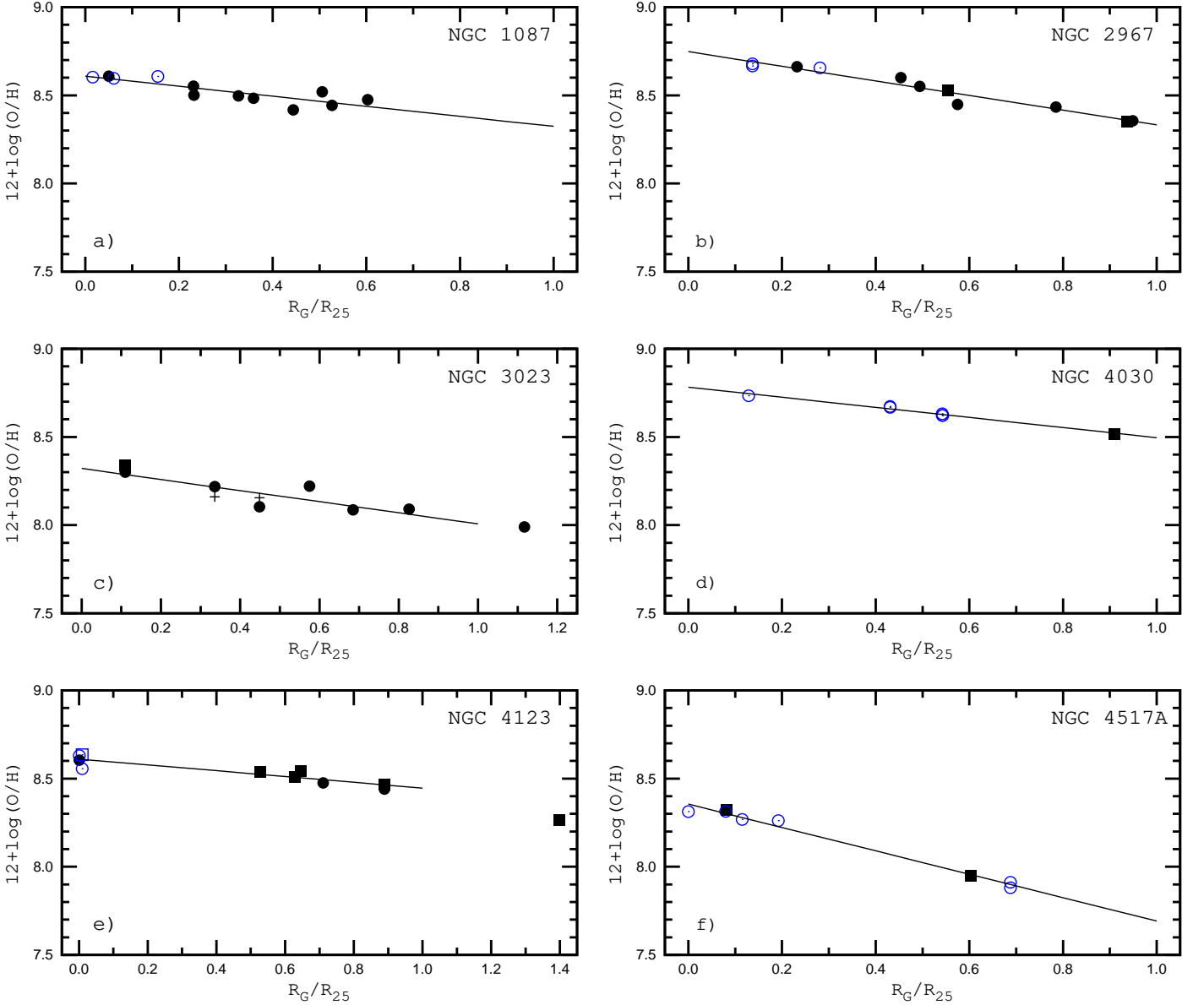}}
\caption{
The radial distributions of oxygen abundances in the disks of our
target galaxies. The plus signs are abundances derived through
the $T_{e}$ method, the circles are abundances obtained through
the $C_{\rm NS}$ method, and the squares are those inferred
through the $C_{\rm ON}$ method. The filled symbols show
abundances based on our SALT spectra, the open (blue) symbols are
abundances based on spectra from the literature (see text).  The solid
line in each panel is the best linear fit to the data points with
galactocentric distances less than the isophotal $R_{25}$ radius.  (A
color version of this figure is available in the online version.) 
}
\label{figure:r-oh}
\end{figure*}

{\bf NGC~1087.}
The strong lines H$\beta$, [O\,{\sc iii}]$\lambda$$\lambda$4959,5007,
H$\alpha$, [N\,{\sc ii}]$\lambda$$\lambda$6548,6584, and [S\,{\sc
ii}]$\lambda$$\lambda$6717,6731 were measured in nine red spectra.  In
those H\,{\sc ii} regions, we derived the oxygen abundance
(O/H)$_{C_{\rm NS}}$.  The resulting oxygen abundances are listed in
Table \ref{table:abundance}.  Those abundances are shown by the filled circles 
in panel $a$ of Fig.~\ref{figure:r-oh}. 

There are three {\it SDSS} spectra of H\,{\sc ii} regions in the
galaxy NGC~1087 in data release 7 \citep[DR7,][]{Abazajian2009}, namely Sp
409-51871-237, Sp 1069-52590-193, and Sp 1511-52946-192.  (The {\it
SDSS} spectrum number consists of the {\it SDSS} plate number, the
modified Julian date of the observation, and the number of the fiber
on the plate.) Since {\it SDSS} data release 10
\citep[DR10,][]{Ahn2014} reported line measurements in one spectrum
only, we used the {\it SDSS} spectra from DR7. The oxygen
(O/H)$_{C_{\rm NS}}$ abundances inferred using the {\it SDSS} spectra
are shown by the open (blue) circles in panel $a$ of
Fig.~\ref{figure:r-oh}. 

The best linear fit to all the data points (12 points) with
galactocentric distances smaller than the isophotal $R_{25}$ radius is  
\begin{equation}
12+\log({\rm O/H})  = 8.61 \pm 0.02 - 0.285 \pm 0.056 \times (R/R_{25}) ,
\label{equation:gN1087}
\end{equation} 
with a mean deviation of 0.034 dex around the relationship. The
obtained relation is shown by a solid line in panel $a$ of
Fig.~\ref{figure:r-oh}.   

{\bf NGC~2967.}
The strong lines H$\beta$, [O\,{\sc iii}]$\lambda$$\lambda$4959,5007,
H$\alpha$, [N\,{\sc ii}]$\lambda$$\lambda$6548,6584 and [S\,{\sc
ii}]$\lambda$$\lambda$6717,6731 were measured in six red spectra.
Oxygen (O/H)$_{C_{\rm NS}}$ abundances were then inferred for those
H\,{\sc ii} regions.  These abundances are shown by the filled
circles in panel $b$ of Fig.~\ref{figure:r-oh}.  The strong lines
[O\,{\sc ii}]$\lambda$$\lambda$3727,3729, H$\beta$, [O\,{\sc
iii}]$\lambda$$\lambda$4959,5007, H$\alpha$, and [N\,{\sc
ii}]$\lambda$6584 were measured in two blue spectra of H\,{\sc ii}
regions in the galaxy  NGC~2967 and were used to estimate oxygen
(O/H)$_{C_{\rm ON}}$ abundances.  Those abundances are shown by the
black filled squares in panel $b$ of Fig.~\ref{figure:r-oh}.  The
obtained  oxygen abundances are given in Table \ref{table:abundance}.

There are three {\it SDSS} spectra (Sp 476-52314-622, Sp
266-51630-387, and 266-51602-394) of H\,{\sc ii} regions in the galaxy
NGC~2967 in DR7.  The oxygen (O/H)$_{C_{\rm NS}}$ abundances derived 
using the {\it SDSS} spectra are shown by the gray (blue) open circles in
panel $b$ of Fig.~\ref{figure:r-oh}. 

The best linear fit to all the data points (11 points) with
galactocentric distances smaller the isophotal $R_{25}$ radius is  
\begin{equation}
12+\log({\rm O/H})  = 8.75 \pm 0.02 - 0.414 \pm 0.030 \times (R/R_{25}) ,
\label{equation:gN2967}
\end{equation} 
with a mean deviation of 0.026 dex around the relationship. The
resulting relation is represented by a solid line in panel $b$ of
Fig.~\ref{figure:r-oh}.   

{\bf NGC~3023.}
The auroral line $R$  =  [O\,{\sc iii}]$\lambda$4363 was detected in
two blue spectra of H\,{\sc ii} regions in the disk of NGC~3023.
The oxygen abundances in those H\,{\sc ii} regions were determined
through the direct $T_{e}$ method. Those abundances are shown by the
plus signs in panel $c$ of Fig.~\ref{figure:r-oh}.
The strong lines [O\,{\sc ii}]$\lambda$$\lambda$3727,3729, H$\beta$,
[O\,{\sc iii}]$\lambda$$\lambda$4959,5007, H$\alpha$, and [N\,{\sc
ii}]$\lambda$6548 were measured in one blue spectrum.  We derived  
the oxygen (O/H)$_{C_{\rm ON}}$ abundance, finding the total nitrogen
flux $N_{2}$ to be  4 [N\,{\sc ii}]$\lambda$6548.  This abundance is
shown by the black filled square in panel $c$ of
Fig.~\ref{figure:r-oh}.  The strong lines H$\beta$, [O\,{\sc
iii}]$\lambda$$\lambda$4959,5007, H$\alpha$, [N\,{\sc
ii}]$\lambda$$\lambda$6548,6584 and [S\,{\sc
ii}]$\lambda$$\lambda$6717,6731 were measured in seven red spectra and
used to infer the oxygen (O/H)$_{C_{\rm NS}}$ abundance for those
H\,{\sc ii} regions.  The total nitrogen fluxes were determined to be
$N_2$  = 1.33[N\,{\sc ii}$\lambda$6584.  Those abundances are shown by
the black filled circles in panel $c$ of Fig.~\ref{figure:r-oh}. 

There are four {\it SDSS} spectra (Sp 480-51989-056, Sp 481-51908-289,
Sp 267-51608-384, and Sp 267-51608-389) of H\,{\sc ii} regions in the
galaxy  NGC~3023.  Since there is a large discrepancy between the line
fluxes reported in DR7 and  DR10 these {\it SDSS} spectra were not used. 
 
The best linear fit to the data points (9 points) with galactocentric
distances smaller than the isophotal $R_{25}$ radius is  
\begin{equation}
12+\log({\rm O/H})  = 8.32 \pm 0.04 - 0.315 \pm 0.078 \times (R/R_{25}) ,
\label{equation:gN3023}
\end{equation} 
with a mean deviation of 0.047 dex around the relationship. The
obtained relation is plotted by a solid line in panel $c$ of
Fig.~\ref{figure:r-oh}.   

{\bf NGC~4030.}
Unfortunately, the spectral setup used for this galaxy covers the
wavelengths of the H$\alpha$ and [N\,{\sc ii}]$\lambda$6584 lines only
for one of the slits. Therefore, the strong lines [O\,{\sc
ii}]$\lambda$$\lambda$3727,3729, H$\beta$, [O\,{\sc
iii}]$\lambda$$\lambda$4959,5007, H$\alpha$, and [N\,{\sc
ii}]$\lambda$6584 were measured in only one blue spectrum of an
H\,{\sc ii} region in the galaxy NGC~4030.  The oxygen (O/H)$_{C_{\rm
ON}}$ abundance was estimated using the measured strong lines.  The
inferred abundance is shown by the black filled squares in panel
$d$ of Fig.~\ref{figure:r-oh}.  There are six {\it SDSS} spectra (Sp
285-51930-042, Sp 285-51663-044, Sp 285-51930-049, Sp 285-51663-058,
Sp 331-52368-405, and Sp 2892-54552-293) of H\,{\sc ii} regions in the
galaxy  NGC~4030 in DR7.  The oxygen (O/H)$_{C_{\rm NS}}$
abundances based on the {\it SDSS} spectra are shown by the gray (blue)
open circles in panel $d$ of Fig.~\ref{figure:r-oh}. 

The best linear fit to all the data points (7 points) with
galactocentric distances smaller the isophotal $R_{25}$ radius is
\begin{equation} 
12+\log({\rm O/H})  = 8.78 \pm 0.01 - 0.286 \pm 0.017 \times (R/R_{25}) , 
\label{equation:gN4030} 
\end{equation} 
with a mean deviation of 0.009 dex from the relationship. The relation
is shown by a solid line in panel $d$ of Fig.~\ref{figure:r-oh}.   

{\bf NGC~4123.}
The strong lines [O\,{\sc ii}]$\lambda$$\lambda$3727,3729, H$\beta$,
[O\,{\sc iii}]$\lambda$$\lambda$4959,5007, H$\alpha$, and [N\,{\sc
ii}]$\lambda$6584 were measured in five blue spectra of H\,{\sc ii}
regions in the galaxy  NGC~4123.  The oxygen (O/H)$_{C_{\rm ON}}$
abundances based on those spectral data are shown by the black
filled squares in panel $e$ of Fig.~\ref{figure:r-oh}.  The sulfur lines
[S\,{\sc ii}]$\lambda$$\lambda$6717,6731 were also
measured in the blue spectrum of one H\,{\sc ii} region in the galaxy
NGC~4123.  This allowed us to obtain the oxygen (O/H)$_{C_{\rm NS}}$
abundance for this particular H\,{\sc ii} region.  The strong lines
H$\beta$, [O\,{\sc iii}]$\lambda$$\lambda$4959,5007, H$\alpha$,
[N\,{\sc ii}]$\lambda$$\lambda$6548,6584 and [S\,{\sc
ii}]$\lambda$$\lambda$6717,6731 were measured in two red spectra.  The
inferred oxygen (O/H)$_{C_{\rm NS}}$ abundances are shown by the 
black filled circles in panel $e$ of Fig.~\ref{figure:r-oh}.  It
should be noted that the H\,{\sc ii} region at the center of NGC~4123
(Slit 06) is located close to the line dividing AGNs and star-forming
regions in the BPT classification diagram. 

Spectra of the region near the center of the NGC~4123 were observed by
\citet{Kehrig2004} and by the {\it SDSS} (Sp 517-52024-504).  We
obtained abundances of 12+log(O/H)$_{C_{NS}}$ = 8.56 and
12+log(O/H)$_{C_{ON}}$ = 8.63 using the spectral measurements of
\citet{Kehrig2004}.  Moreover, we measured an abundance of
12+log(O/H)$_{C_{NS}}$ = 8.63 using the DR10 line fluxes. 

The best linear fit to all the data points (10 points) with
galactocentric distances smaller than the isophotal $R_{25}$ radius is  
\begin{equation}
12+\log({\rm O/H})  = 8.61 \pm 0.01 - 0.164 \pm 0.025 \times (R/R_{25}) ,
\label{equation:gN4123}
\end{equation} 
with a mean deviation of 0.025 dex. This relation is represented by a
solid line in panel $e$ of Fig.~\ref{figure:r-oh}.   

{\bf NGC~4517A.}
The strong lines [O\,{\sc ii}]$\lambda$$\lambda$3727,3729, H$\beta$,
[O\,{\sc iii}]$\lambda$$\lambda$4959,5007, H$\alpha$, and [N\,{\sc
ii}]$\lambda$6584 were measured in the blue spectrum of an H\,{\sc ii}
region in the galaxy  NGC~4517A.  In another spectrum, the line
[N\,{\sc ii}]$\lambda$6584 is out of our spectral range, but the line
[N\,{\sc ii}]$\lambda$6548 is included.  We derived the oxygen
(O/H)$_{C_{\rm ON}}$ abundance in these spectra.  The total
nitrogen flux $N_{2}$ was determined to be $N_2$  = 1.33[N\,{\sc
ii}]$\lambda$6584 in the former case and to be $N_2$  = 4 [N\,{\sc
ii}]$\lambda$6548 in the latter case.  These abundances are shown by
the black filled squares in panel $f$ of Fig.~\ref{figure:r-oh}. 

\citet{Romanishin1983} reported emission-line ratios [S\,{\sc
ii}]($\lambda 6717+\lambda 6731$)/H$\alpha$, H$\alpha$/[N\,{\sc
ii}]($\lambda 6548+\lambda 6584$), and [O\,{\sc
iii}]($\lambda$4959+$\lambda$5007)/H$\beta$ obtained from photographic
spectra of four H\,{\sc ii} regions in NGC~4517A. We estimated the
oxygen (O/H)$_{C_{\rm SN}}$ and nitrogen (N/H)$_{C_{\rm SN}}$
abundances from those strong lines.  \citet{Romanishin1983} did not
provide the positions of the observed H\,{\sc ii} regions, but listed
the deprojected radii instead.  We corrected these galactocentric
distances for the galaxy distance adopted here and used the resulting
values.  Furthermore, there are two {\it SDSS} spectra (Sp
289-51990-627 and Sp 290-51941-350) of H\,{\sc ii} regions in
NGC~4517A.  The abundances based on the {\it SDSS} and Romanishin et
al.'s data are shown by the gray (blue) open circles in panel $f$ of
Fig.~\ref{figure:r-oh}. 

The best linear fit to all the data points (8 points) with
galactocentric distances smaller than the isophotal $R_{25}$ radius is  
\begin{equation}
12+\log({\rm O/H})  = 8.35 \pm 0.01 - 0.663 \pm 0.033 \times (R/R_{25}) ,
\label{equation:gN4517}
\end{equation} 
with a mean deviation of 0.023 dex. This relation is indicated by a
solid line in panel $f$ of Fig.~\ref{figure:r-oh}.   

\begin{figure*}
\resizebox{1.00\hsize}{!}{\includegraphics[angle=000]{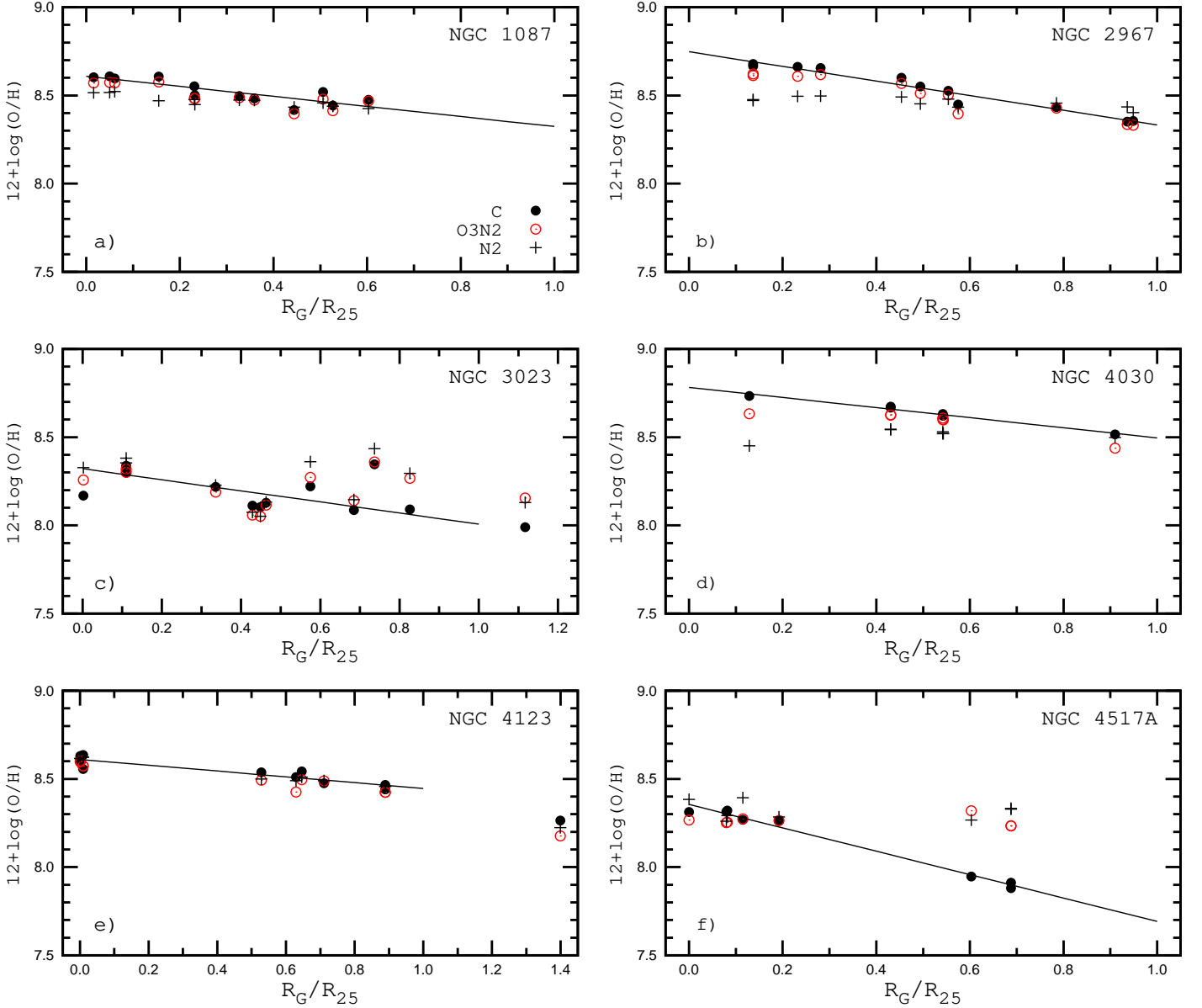}}
\caption{
Comparison of the radial distributions of oxygen abundances in the disks of our
target galaxies determined through the $C$ method (filled dark (black) circles), 
via the $O3N2$ calibration (open grey (red) circles), and through 
the $N2$ calibration (plus signs). 
The solid line in each panel is the best linear fit to the (O/H)$_{C}$ 
abundances (the same as in  Fig.~\ref{figure:r-oh}). 
(A color version of this figure is available in the online version.) 
}
\label{figure:r-ohmar}
\end{figure*}

\subsection{Comparison between the distributions of the abundances 
determined through the $C$ method and via the O3N2 and N2 calibrations}

Many calibrations based on photoionization models or/and  H\,{\sc ii} regions 
with abundances determined through the direct $T_{e}$-method were suggested for 
nebular abundance determinations \citep[][among many others]{Pagel1979,Alloin1979,Dopita1986,McGaugh1991,
Zaritsky1994,Pilyugin2000,Pilyugin2001,Kewley2002,Pettini2004,Pilyugin2005,Tremonti2004,Stasinska2006}.
The discrepancies between metallicities of a given  H\,{\sc ii} region derived 
using different calibrations can be large, up to $\sim$0.6 dex \citep[see reviews
by][]{Kewley2008,LopezSanchezEsteban2010AA517,LopezSanchezetal2012MNRAS426}. 
However, one would expect that all the calibrations based on the abundances in 
 H\,{\sc ii} regions determined through the $T_{e}$ method should produce 
the abundances close to each other.  
Since the $C$ method produces abundances on the same metallicity scale as the $T_{e}$-method 
those abundances should be close to the abundances produced by other 
calibrations based on the direct abundances in  H\,{\sc ii} regions. 
 
The O3N2 and N2 calibrations suggested by \citet{Pettini2004} are widely used. The original 
O3N2 and N2 calibrations are hybrid calibrations, i.e., they are based on both  H\,{\sc ii} regions 
with abundances determined  through the direct $T_{e}$-method and photoionization models. 
``Pure'' empirical O3N2 and N2 calibrations (i.e., based on H\,{\sc ii} regions 
with abundances determined  through the direct $T_{e}$-method) were recently presented 
by \citet{Marino2013}. 
Thus we determined the (O/H)$_{O3N2}$ abundances in our  H\,{\sc ii} regions also using the 
O3N2 calibration of \citet{Marino2013}
\begin{equation}
12+\log({\rm O/H})_{O3N2}  = 8.533  - 0.214 \times O3N2 
\label{equation:oho3n2}
\end{equation} 
where O3N2 = log[(${\rm [O\,III]}\lambda 5007/{\rm H}\beta)$/${\rm [N\,II]}\lambda  6584/{\rm H}\alpha)]$, 
and the (O/H)$_{N2}$ abundances  using their N2 calibration 
\begin{equation}
12+\log({\rm O/H})_{N2}  = 8.743  + 0.462 \times N2 
\label{equation:oho3n2}
\end{equation} 
where N2 = log(${\rm [N\,II]\lambda  6584}/{\rm H}\alpha$). 

Fig.~\ref{figure:r-ohmar} shows the comparison of the radial distributions of the oxygen abundances 
in the disks of our
target galaxies determined through the $C$ method (filled dark (black) circles), 
via the $O3N2$ calibration (open grey (red) circles), and through 
the $N2$ calibration (plus signs). 
The solid line in each panel is the best linear fit to the (O/H)$_{C}$ 
abundances (the same as in  Fig.~\ref{figure:r-oh}).

\begin{figure}
\resizebox{0.75\hsize}{!}{\includegraphics[angle=000]{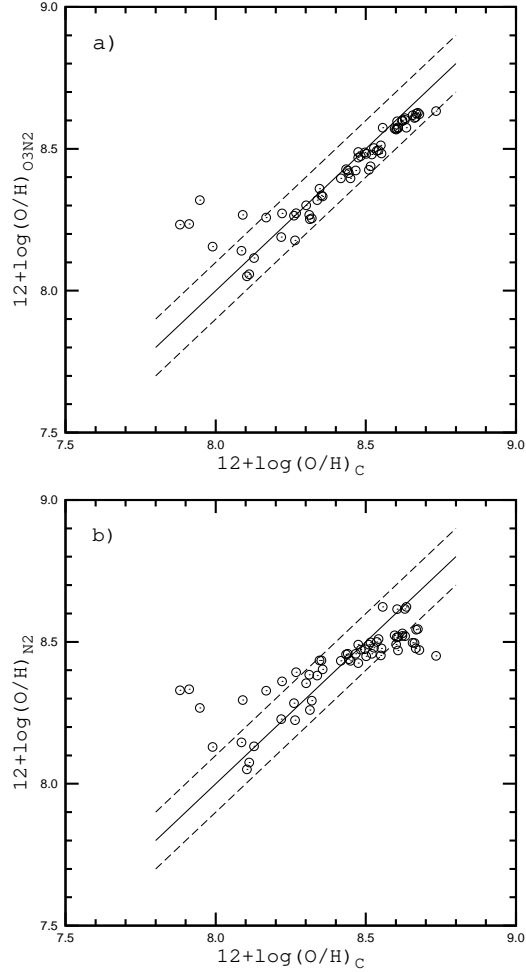}}
\caption{
Comparison of the oxygen abundances in the individual H\,{\sc ii} regions of our
sample determined through the $C$ method with oxygen abundances obtained through 
the $O3N2$ calibration (upper panel) and through 
the $N2$ calibration (lower panel). 
The solid line indicates a one-to-one correspondence. The dashed lines are shifted by $\pm 0.1$ dex.
}
\label{figure:ohc-ohmar}
\end{figure}

Fig.~\ref{figure:ohc-ohmar} shows the comparison of the oxygen abundances in the individual H\,{\sc ii} regions of our
sample determined through the $C$ method with the oxygen abundances obtained through 
the $O3N2$ calibration (upper panel) and through 
the $N2$ calibration (lower panel). 

Fig.~\ref{figure:r-ohmar} and Fig.~\ref{figure:ohc-ohmar} demonstrate that the (O/H)$_{O3N2}$ 
abundances are in satisfactory agreement (within 0.1 dex) with the (O/H)$_{C}$ abundances for H\,{\sc ii} 
regions with metallicities 12+log(O/H) $\ga$ 8.1. 
However, a small systematic difference (around 0.05 dex) between (O/H)$_{O3N2}$ and (O/H)$_{C}$ abundances 
seems to exist; in the sense that the (O/H)$_{O3N2}$ abundances are slightly lower than the (O/H)$_{C}$ abundances. 
A large disagreement between (O/H)$_{O3N2}$ and (O/H)$_{C}$ abundances for H\,{\sc ii} 
regions with metallicities 12+log(O/H)$_{C}$ $\la$ 8.1 is not surprising since the O3N2 calibration 
of \citet{Marino2013} is constructed for H\,{\sc ii} regions with metallicities 12+log(O/H) $\ga$ 8.1 
and does not work at low metallicities. 
The differences between the (O/H)$_{N2}$ and (O/H)$_{C}$ abundances exceed 0.1 dex for 
some H\,{\sc ii} regions with metallicities 12+log(O/H) $\ga$ 8.1. 
This may suggest that the O3N2 calibration of \citet{Marino2013} provides more reliable abundances 
than their N2 calibration.   

In summary, the comparison between $C$-,  O3N2-, and N2-based abundances in our target H\,{\sc ii} regions 
allows us to suggest that the uncertainties in the obtained (O/H)$_{C}$ abundances are within 
$\sim 0.1$ dex. This supports our estimation of the uncertainties in the abundances discussed 
in Subsection 5.2.

\section{Discussion}

\begin{figure*}
\resizebox{1.00\hsize}{!}{\includegraphics[angle=000]{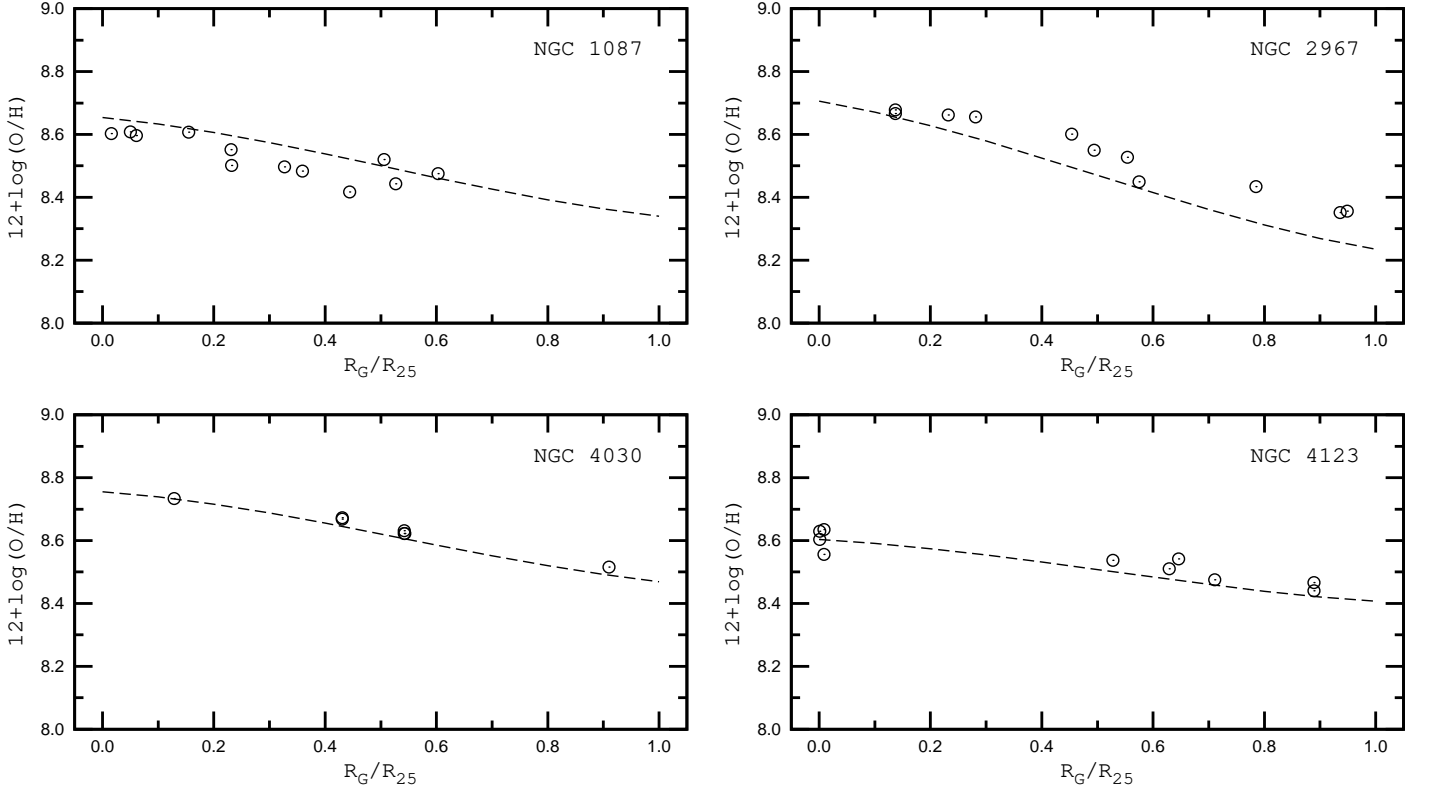}}
\caption{
Radial distributions of oxygen abundances in the disks of the spiral
galaxies of our sample.  The circles represent the abundances of the
individual H\,{\sc ii} regions (the same as in
Fig.~\ref{figure:r-oh}).  The line in each panel shows the abundance
distribution predicted by the relation between abundance and surface
brightness in the $W1$ band,  Eq.~(\ref{equation:oh-sb}). 
}
\label{figure:oh-sb}
\end{figure*}

The radial distributions of the oxygen abundances across the disks of
all the galaxies of our sample are well fitted by linear relationships
within the isophotal radius (with the abundances on the logarithmic
scale).  The mean deviation from the relationship is less than 0.05
dex for each galaxy. The values of the radial abundance gradient vary
by a factor of $\sim 4$ among the galaxies of our sample; from
$-0.164$~dex $R_{25}^{-1}$ for NGC~4123 to $-0.663$~dex $R_{25}^{-1}$
for NGC~4517A. 

The correlation between the local oxygen abundance and the stellar
surface brightness (OH -- $SB$ relation) or surface mass density has
been discussed in many studies
\citep{Webster1983,Edmunds1984,VilaCostas1992,Ryder1995,Moran2012,Rosales2012,Sanchez2014}.
In our previous paper \citep{Pilyugin2014b} we examined the relations
between the oxygen abundance and the disk surface brightness in the
infrared $W1$ band of {\it WISE} at different fractions of the optical 
isophotal radius
$R_{25}$.  We found evidence that the OH -- $SB$ relation varies with
galactocentric distance and depends on the disk scale length and
the morphological $T$-type of a galaxy.  We derived a general
parametric relation between abundance and surface brightness in the
$W1$ band, O/H = $f(SB)$ for spiral galaxies of type $Sa$ -- $Sd$, 
\begin{eqnarray}
       \begin{array}{lll}
12+\log {\rm (O/H)}    & = &  7.732 + 0.303 \, x + 0.290 \, x^{2}  \\  
                       & + & (0.288 + 0.120 \, x \\ 
                       & & \,\,  - 0.139 \, x^{2}) \, \log (\Sigma_{L_{W1}})_{x} \\  
                       & + & (0.0418 - 0.0323 \, x - 0.0022 \, x^{2}) \, h_{W1}  \\
                       & - & (0.0122 + 0.0404 \, x + 0.0088 \, x^{2}) \, T \\
     \end{array}
\label{equation:oh-sb}
\end{eqnarray}
where $x$ = $r/R_{25}$ is the fractional radius expressed in terms of
the isophotal radius of a galaxy ($R_{25}$), $(\Sigma_{L})_{x}$ is the disk
surface brightness, $h_{W1}$ the radial disk scale length, and $T$ the
morphological $T$-type.  It is interesting to compare the radial
distributions of oxygen abundances predicted by this relationship to
the radial abundance trends traced by the oxygen abundances in the
H\,{\sc ii} regions in the disks of our sample of galaxies. 

Fig.~\ref{figure:oh-sb} shows the comparison between the radial
distributions of the oxygen abundances predicted by the O/H = $f(SB)$
relation,  Eq.~(\ref{equation:oh-sb}), and the abundances obtained
from the analysis of the emission-line spectra of H\,{\sc ii} regions
for four of our galaxies, NGC~1087, NGC~2967, NGC~4030, and NGC~4123.
The O/H = $f(SB)$ relation cannot be applied to the other two galaxies
of our sample. It was noted above that we could not determine a
reliable disk scale length $h_{W1}$ and surface brightness at the
center of the disk of the galaxy NGC~3023 since the disk contribution
to the surface brightness is close to the observed surface-brightness
profile over a small range of radial distances only (see lower panel
of Figure~\ref{figure:decomposit}).  NGC~4517A is a $Sdm$ galaxy (with
morphological type $T$ = 8) whereas the O/H = $f(SB)$ relation was
derived for spiral galaxies of the types $Sa$ -- $Sd$ (i.e., for a
range of morphological types from $T \sim 1$ to $T \sim 7$).

Inspection of Fig.~\ref{figure:oh-sb} shows that the oxygen abundances
predicted by the parametric O/H = $f(SB)$ relation are rather close to
the abundances obtained from the analysis of the emission-line spectra
of  H\,{\sc ii} regions of the galaxies of the present sample where the
OH -- $SB$ relation is applicable. The discrepancy usually does not
exceed 0.1 dex. Thus, the parametric O/H = $f(SB)$ relation can be
used for a rough estimation of the oxygen abundances in the disks of
spiral galaxies.

\section*{Summary} 

Spectra of H\,{\sc ii} regions in six late-type galaxies were observed
with the South African Large Telescope (SALT). The auroral line
[O\,{\sc iii}]$\lambda$4363 was detected in two spectra. The
$T_{e}$-based oxygen (O/H)$_{T_{e}}$ abundances in these two H\,{\sc ii}
regions were derived using the equations of the standard
$T_{e}$-method. The oxygen abundances of the other H\,{\sc ii} regions
were estimated from strong emission lines through the recently
suggested ``counterpart'' method ($C$ method). When the strong lines
$R_3$, $N_2$, and $S_2$ were measured in our spectra, oxygen
(O/H)$_{C_{\rm NS}}$ abundances could be obtained.  When, on the other
hand, the strong lines $R_2$, $R_3$, and $N_2$ were available then
oxygen (O/H)$_{C_{\rm ON}}$ abundances were determined.  Moreover, we
also inferred oxygen abundances of the H\,{\sc ii} regions in our
target galaxies with available spectral measurements from the
literature or from the {\it SDSS} spectroscopic data base through the
$C$ method.

We derived oxygen abundances from the individual one-pixel-wide
spectra and considered the variations in those abundances. The
abundances determined with the $C$ method from the individual
one-pixel-wide spectra are close to each other and are close to the
abundances obtained from the integrated seven-pixel-wide spectrum.
This can be considered as supporting evidence for the robustness and
precision of the $C$-based abundances, which are independent of the
area in the H\,{\sc ii} region image that is measured. In other words,
the $C$ method produces a reliable oxygen abundance even if a spectrum
of only a part of an H\,{\sc ii} region is used.  

We also determined the (O/H)$_{O3N2}$ and (O/H)$_{N2}$  abundances in our target  
H\,{\sc ii} regions using the O3N2 and N2 calibrations of \citet{Marino2013}. 
The (O/H)$_{O3N2}$ abundances are in satisfactory agreement (within 0.1 dex) 
with the (O/H)$_{C}$ abundances for H\,{\sc ii} regions with metallicities 
12+log(O/H) $\ga$ 8.1. However, a small systematic difference (around 0.05 dex) 
between (O/H)$_{O3N2}$ and (O/H)$_{C}$ abundances seems to exist in the sense 
that the (O/H)$_{O3N2}$ abundances are slightly lower than the (O/H)$_{C}$ abundances. 
The differences between the (O/H)$_{N2}$ and (O/H)$_{C}$ abundances are larger than 
0.1 dex for some H\,{\sc ii} regions. This may suggest that the O3N2 calibration 
of \citet{Marino2013} provides more reliable abundances than their N2 calibration.

We determined the abundance gradients in the disks of our six
late-type target galaxies.  The radial distributions of the oxygen
abundances across the disks of all the galaxies of our sample are well
fitted by linear relationships within the isophotal radius (with
abundances on a logarithmic scale).  The mean deviation from the
relationship is less than 0.05 dex for each galaxy. The values of the
radial abundance gradient vary by a factor of $\sim 4$ among the
galaxies of our sample, i.e., from $-0.164$~dex $R_{25}^{-1}$ for
NGC~4123 to $-0.663$~dex $R_{25}^{-1}$ for NGC~4517A. 

We derived surface-brightness profiles in three photometric bands (the
$W1$ band of {\it WISE} and the $g$ and $r$ bands of the {\it SDSS})
for each galaxy using publicly available photometric imaging data. The
characteristics of the disks (the surface brightness at the disk
center and the disk scale length) were found through bulge-disk
decomposition.  Using the photometric parameters of the disks, the
oxygen abundance distributions were estimated from the relation
between abundance and surface brightness of the disk in the $W1$ band,
O/H = $SB$, which had been obtained for spiral galaxies in our
previous study. The oxygen abundances predicted by the O/H = $SB$
relation are rather close to the abundances determined from the
analysis of the emission-line spectra of the H\,{\sc ii} regions in
the galaxies of the present sample where the OH -- $SB$ relation is
applicable. The discrepancy is usually not larger than 0.1 dex. Thus,
the parametric O/H = $f(SB)$ relation can be used for a rough
estimation of the oxygen abundances in the disks of spiral galaxies.  

\section*{Acknowledgements}

The observations reported in this paper were obtained with the
Southern African Large Telescope (SALT).  L.S.P., I.A.Z., and E.K.G.\
acknowledge support within the framework of Sonderforschungsbereich
(SFB 881) on ``The Milky Way System'' (especially subproject A5),
which is funded by the German Research Foundation (DFG).  L.S.P.\ and
I.A.Z.\ thank the Astronomisches Rechen-Institut at the
Universit\"{a}t Heidelberg  where this investigation was carried out
for the hospitality.  A.Y.K.\ acknowledges the support from the
National Research Foundation (NRF) of South Africa.
This work was partly funded by the subsidy allocated to the Kazan Federal 
University for the state assignment in the sphere of scientific 
activities (L.S.P.).

The authors
acknowledge the work of the {\it SDSS} collaboration.  Funding for SDSS-III
has been provided by the Alfred P.\ Sloan Foundation, the
Participating Institutions, the National Science Foundation, and the
U.S.\ Department of Energy Office of Science. The SDSS-III web site is
http://www.sdss3.org/.  SDSS-III is managed by the Astrophysical
Research Consortium for the Participating Institutions of the SDSS-III
Collaboration including the University of Arizona, the Brazilian
Participation Group, Brookhaven National Laboratory, University of
Cambridge, Carnegie Mellon University, University of Florida, the
French Participation Group, the German Participation Group, Harvard
University, the Instituto de Astrofisica de Canarias, the Michigan
State/Notre Dame/JINA Participation Group, Johns Hopkins University,
Lawrence Berkeley National Laboratory, Max Planck Institute for
Astrophysics, Max Planck Institute for Extraterrestrial Physics, New
Mexico State University, New York University, Ohio State University,
Pennsylvania State University, University of Portsmouth, Princeton
University, the Spanish Participation Group, University of Tokyo,
University of Utah, Vanderbilt University, University of Virginia,
University of Washington, and Yale University.\\ We acknowledge the
usage of the HyperLeda database (http://leda.univ-lyon1.fr).

\bibliography{SALT-final.bib}

\end{document}